\documentclass[aps,prb,nopacs,floatfix,12pt]{revtex4}

\usepackage{graphicx}
\usepackage{amssymb}
\usepackage{amsmath}
\usepackage{epsfig}
\usepackage{bm}
\graphicspath{{/home/sgreben/figures/no2-kin/}}

\begin{document}
\bibliographystyle{jcp}

\title{Photodissociation of carbon dioxide in singlet valence
electronic states. 
I. Six multiply intersecting ab initio potential energy surfaces}

\author{Sergy\ Yu.\ Grebenshchikov\footnote{Sergy.Grebenshchikov@ch.tum.de}}
\affiliation{Department of Chemistry, Technical University of Munich,
  Lichtenbergstr. 4, 85747 Garching, Germany} 

\begin{abstract}
The global potential energy surfaces of the first six singlet
electronic states of CO$_2$, 1---3$^1\!A'$ and 1---3$^1\!A"$ are
constructed using high level ab initio calculations. 
In linear molecule, they correspond to $\tilde{X}^1\Sigma_g^+$, 
$1^1\Delta_u$, $1^1\Sigma_u^-$, and $1^1\Pi_g$. The calculations
accurately reproduce the known benchmarks for all states and establish
missing benchmarks for future calculations. The calculated states
strongly interact at avoided crossings and true intersections, both
conical and glancing. Near degeneracies can be found for each pair of
six states and many intersections involve more than two states. In
particular, a fivefold intersection dominates the Franck-Condon zone
for the ultraviolet excitation from the ground electronic state. The
seam of this intersection traces out a closed loop. All states are
diabatized, and a diabatic $5\times 5$ potential matrix is constructed,
which can be used in  
quantum mechanical calculations of the absorption spectrum
of the five excited singlet valence states. 

\end{abstract}

\maketitle

\section{Introduction}\label{introduction}

This and the subsequent\cite{G13B} paper (termed \lq paper II')
describe the results of an 
ab initio quantum dynamical study of the absorption spectrum and the 
non-adiabatic dissociation mechanisms of carbon dioxide photoexcited with the
ultraviolet (UV) light between 120\,nm and 160\,nm. A brief account of this 
work has already been published.\cite{G12A}
Paper II gives the motivation behind the study and
 discusses its main result --- the quantum mechanical absorption spectrum
and its interpretation in terms of  
wave functions of metastable resonance states. The present paper 
sets the stage for paper II and describes the 
ab initio calculations and the topography of the potential energy
surfaces (PESs) 
involved in photodissociation. This information, which is often stashed away 
in supplementary online sections,\cite{G12A,GB12} is a central 
and, indeed, an indispensable ingredient of a reliable dynamics
calculation. The
construction of ab initio PESs and their diabatization --- which,
without much exaggeration,  amounts to learning 
the topography of PESs and their intersections by heart --- 
is often more challenging than the subsequent quantum dynamical calculation.

Photoabsorption from the ground electronic state $\tilde{X}^1\Sigma_g^+$
of linear CO$_2$ 
at wavelengths 120\,nm --- 160\,nm is due to the first five
excited singlet valence states $1^1\Sigma_u^-$,  
 $1^1\Pi_g$, and $1^1\Delta_u$.\cite{HERZBERG67,RMSM71,OKABE78} 
In the $C_s$ group notation, appropriate for bent molecule, 
these states are $2,3^1\!A'$ 
and $1,2,3^1\!A''$.  UV light excites CO$_2$ into the region
of multiple electronic degeneracies, 
nuclear motion through which induces strong non-adiabatic couplings between
electronic states. 
These couplings directly affect the observed absorption spectrum of the
valence states and control
distributions of photofragments over the final states. Their indirect
influence apparently extends to shorter wavelengths where  
Rydberg transitions dominate: The combined experimental and theoretical analysis
indicates\cite{CJL87} that the manifold of coupled valence states acts as a \lq
sink' for the optically bright Rydberg states and affects their dissociation
lifetimes.  

Electronic degeneracies in the Franck-Condon (FC) region, which  have been the
focus of several studies  in the past,\cite{EE79,KRW88,SFCCRWB92,G12A,GB12} 
are of two types. Glancing intersections occur in the orbitally doubly
degenerate $1^1\Pi_g$ and $1^1\Delta_u$ states which upon bending 
split into $A'$ and $A''$ components.  Conical intersections
(CIs) arise at the accidental  $1^1\Pi_g/1^1\Delta_u$ 
crossing inside $A'$ and $A''$ symmetry blocks.\cite{KRW88,SFCCRWB92,G12A,GB12}
In fact, CIs  between valence states are ubiquitous and found far outside
the near-linear FC region, at strongly bent 
geometries.\cite{SFCCRWB92,G12A,GB12} Together with local 
minima and saddles,  
these crossings are the principal features shaping the topography
of the singlet valence states. 

The outline of the paper is the following. Section \ref{abi} sketches the
technical details of ab initio calculations. Next, the constructed adiabatic
PESs are presented for the ground (Sect.\ \ref{gz}) and the 
excited (Sect.\ \ref{ex1}) electronic states.  Their intersections are discussed
for near linear (Sect.\ \ref{ex2}) 
and bent geometries (Sect.\ \ref{ex4}), and put into perspective by a
review of a network 
of closely spaced valence and Rydberg states (Sect.\ \ref{ex5}). 

If the solution of the Schr\"odinger equation for nuclei is out of reach,
the description of adiabatic surfaces in Sect.\ \ref{topi} would be the
last step in theoretical ab initio analysis. If, on the other
hand, one intends to treat nuclear dynamics quantum mechanically, 
ab initio PESs featuring CIs 
have to be diabatized.\cite{C04} This is especially
desirable if --- as in paper II --- a discrete grid is used to
represent the nuclear Hamiltonian, because the diabatized potential
matrix is free from either divergent off-diagonal couplings or
non-differentiable potential cusps. While general schemes for 
constructing approximate adiabatic-to-diabatic 
transformations are established (see, for instance, Ref.\
\onlinecite{K04}), their application to the valence states of CO$_2$
is complicated by the number of CIs to be simultaneously
treated. Simplifications are called for, as described in Sect.\
\ref{diabats}, in which the 
diabatic representation is constructed separately 
for bent CIs (Sect.\ \ref{qd1}) and
linear CIs (Sect.\ \ref{qd2}). Section \ref{concl} concludes.

\section{Electronic structure calculations}
\label{abi}

All ab initio calculations are carried
out with the MOLPRO package.\cite{MOLPRO-FULL}
The Gaussian atomic basis sets used in this work are due to 
Dunning.\cite{D89} Previous studies indicate
that diffuse functions should be added to the basis
sets on oxygen and carbon atoms in order to account for the
mixed valence-Rydberg character of the 
$\Pi$ state.\cite{KRW88,SFCCRWB92,G12A} A series of
tests was conducted, in which $s$, $p$, $d$ etc. basis functions
of triple and quadrupole zeta quality were selectively 
augmented by one or two diffuse functions. The pre-computed 
doubly augmented correlation consistent polarized valence quadrupole
zeta (d-aug-cc-pVQZ) basis set,\cite{D89} as implemented in
MOLPRO, was found computationally most stable and selected for 
calculations of global PESs.

Three-dimensional (3D) PESs of states
$1,2,3^1\!A'$ and $1,2,3^1\!A''$ are calculated at the 
internally-contracted multireference configuration interaction singles
and doubles (MRD-CI) level,  based on
state-averaged full-valence complete active space self-consistent
field (CASSCF) calculations 
with 16 electrons in 12 active orbitals and 6 electrons
in three fully optimized closed-shell inner orbitals. 
The electronic configuration of the ground state 
$\tilde{X}^1\Sigma_g^+$ is 
$(\mbox{[core]}2\sigma_u^23\sigma_g^24\sigma_g^23\sigma_u^21\pi_u^41\pi_g^4)$.
The dominant electronic excitations, leading to the lowest excited states,
include $1\pi_g^4 \rightarrow 2p2\pi_u^1$ (giving states $1^1\Delta_u$
and $1^1\Sigma_u^-$) and $1\pi_g^4 \rightarrow 5s5\sigma_g^1$ 
(giving state $1^1\Pi_g$). 
Active orbitals in CASSCF comprised 
$2\sigma_u - 4\sigma_u$, $3\sigma_g-5\sigma_g$,$1\pi_u-2\pi_u$, 
and $1\pi_g$. In $C_s$ symmetry, used in the calculations, these are
$4a'- 12a'$ and $1a'' - 3a''$. In the
MRD-CI step all 16 valence electrons were correlated. The
maximum numbers 
of open shells allowed in the MRD-CI calculations were 8 in the reference
space and 12 in the internal space. This lead to 38159928 contracted
configurations. The Davidson correction was 
applied in order to account for higher-level
excitations and size-extensivity.\cite{LD74} 

Adiabatic energies are calculated on a 3D grid of the two
C--O bond lengths $R_{1,2}$ and the OCO bond angle $\alpha_{\rm OCO}$: 
$R_{2} \in [1.9\,a_0,3.0\,a_0]$ (step size equals 0.1\,$a_0$),
$R_{1} \in [R_2,6.2\,a_0]$ (step size varies between 0.1\,$a_0$ and
0.4\,$a_0$),  
$\alpha_{\rm OCO} \in [70^\circ,179.9^\circ]$ (step size varies between 
2$^\circ$ and 10$^\circ$). Additionally, many cuts in the $(R_1,R_2)$ plane are
computed for angles $\alpha_{\rm OCO}$ between 60$^\circ$ and
0$^\circ$ within the 
continuing effort to construct a balanced description of both CO + O
and C + O$_2$ 
arrangement channels. At present, the grid comprises 4800 
symmetry distinguishable points. The resulting energies were
scanned in one and two dimensions 
for obvious errors. The list of corrected adiabatic energies
was subsequently interpolated using 3D cubic splines and also
used for constructing the quasi-diabatic representation. 
Missing energies for $\alpha_{\rm OCO} < 70^\circ$ in the dissociation channels 
were obtained from those for $\alpha_{\rm OCO} > 70^\circ$ using trigonometric 
extrapolation. 

Absolute intensity calculations of paper II require transition dipole moments 
(TDMs) with the ground state $\tilde{X}$. 
Components $(\mu_x,\mu_y,\mu_z)$ of the TDM vector are calculated, for each
electronic state, on a 3D grid $R_{1,2} = [1.9\,a_0,2.4\,a_0]$ and 
$\alpha = [165^\circ - 179^\circ]$ covering the spot
over which the vibrational ground state in $\tilde{X}$ is
delocalized. The molecular axes in these calculations are chosen such
that $x'$ is orthogonal to the molecular plane, $z'$  runs
along one of the CO bonds and $y' \perp z'$. 
For $A'$ states, the
in-plane components are generally non-zero, while for $A''$ states, it
is the $\mu_{x'}$ component which carries the transition. 

\section{Properties of the valence PES$\bf s$ and their crossings}
\label{topi}

\subsection{Ground electronic state}
\label{gz}

The $\sim 7.5$\,eV deep adiabatic
ground state PES supports three structural isomers: 
The familiar linear OCO molecule is the global equilibrium, while 
the carbene-like bent OCO and the linear COO are the two local ones. 
Table \ref{table1}
summarizes the characteristic features of the $\tilde{X}^1\Sigma_g^+$ state
at the three equilibria and 
compares them with the previous ab initio studies and
with the available experimental data. 

The vicinity of the global minimum is of capital importance for the
environmental chemistry.\cite{P11}  
The calculated equilibrium CO bond distance in linear OCO, 
$R_{e} = 2.1991\,a_0$, agrees well with the
experimental value of 2.1960\,$a_0$. The accuracy of the vibrational
zero-point energy (ZPE) and the vibrational transitions frequencies
is assessed in Table \ref{table2} which compares energies of the low lying  
vibrationally excited states in rotating CO$_2$ (the total angular momentum 
$N_{\rm CO2} \ge 0$) with experiment\cite{C79A} and with recent
electronic coupled cluster/vibrational configuration interaction 
calculations.\cite{RHYHTST07,YHH08} 
Each eigenstate $(v_s,v_b^l,v_a)$ is labeled using the quantum numbers 
of the symmetric stretch $v_s$, the bend $v_b^l$ (with $l$ indicating the 
vibrational angular momentum, $N_{\rm CO2} \ge l$),
 and the antisymmetric stretch $v_a$. The calculated fundamental 
frequencies of the infrared active bend 
($\omega_{b} = 668.6$\,cm$^{-1}$)
and antisymmetric stretch  ($\omega_{a} = 2350.6$\,cm$^{-1}$) 
are accurate to within 1.5\,cm$^{-1}$. 
The zeroth order symmetric stretch frequency\cite{NISTDATABASE1} 
$\omega_{s}^0 \approx 1333$\,cm$^{-1}$ is about twice as large as the bending 
frequency $\omega_{b}$, 
and the two modes are involved in the accidental Fermi resonance.\cite{F31}
As a result, the vibrational spectrum is organized in 
polyads with the polyad quantum number $P = 2v_s+v_b$; states with
$P = 2$, 3 and 4 are given in Table \ref{table2}.   In the original version
of the PES, called \lq PES1' in Table \ref{table2}, the energies of states 
$(1,0^0,0)$ and $(0,2^0,0)$, belonging to the lowest polyad $P = 2$, 
are underestimated by 20\,cm$^{-1}$ and the difference with the
observed energies grows rapidly with $P$. This systematic discrepancy
is substantially diminished by slightly
 rescaling the symmetric stretch, $R_+ = (R_1+R_2)/\sqrt{2}$,
and the bend $\alpha_{\rm OCO}$ via
\begin{eqnarray*}
\label{rescal}
R_+ & \rightarrow & \sqrt{2}R_{1e} + (R_+ - \sqrt{2}R_{1e})\cdot 1.023\\ 
\alpha_{\rm OCO} & \rightarrow & 180^\circ + (\alpha_{\rm
  OCO}-180^\circ)\cdot1.0035 \, .
\end{eqnarray*}
The vibrational energies in the scaled \lq PES2' agree with their
experimental counterparts to within 7\,cm$^{-1}$ and for the most states
below 3000\,cm$^{-1}$ the accuracy is better than 3\,cm$^{-1}$. The results
outperform even the highly accurate calculations
of Refs.\ \onlinecite{RHYHTST07} and \onlinecite{YHH08} 
shown in the second column of Table
\ref{table2}, making \lq PES2' one of the best available ab initio
potentials of the $\tilde{X}^1\Sigma_g^+$ state. Since the coordinate
dependent dipole moment $\mu_{\tilde{X}}$ has also been calculated,
the ab initio intensities of the infrared rovibrational
transitions can be directly evaluated.\cite{G13C}

The other two isomers in Table \ref{table1} 
have never been detected in the gas phase, and
the only reference data stem from the previous  ab initio studies. 
For the bent OCO, discovered by Xantheas and Ruedenberg,\cite{XR94} the
present calculations confirm the $C_{2v}$ symmetric equilibrium
with the CO bond lengths of 2.51\,$a_0$ and the OCO bond
angle of 73.2$^\circ$. This minimum is located 
6.03\,eV above the global one, again in good agreement with the previous 
findings.\cite{XR94,HM00A} The fundamental excitations in
the OCO well, calculated for $N_{\rm CO2} = 0$, are 
$\omega_a = 680$\,cm$^{-1}$, $\omega_b = 720$\,cm$^{-1}$, and
$\omega_s = 1550$\,cm$^{-1}$. For the linear COO,  
the calculated CO and OO bond lengths are identical to the ones given 
in Ref.\ \onlinecite{HM00A}; both 
are elongated compared to free diatoms ($2.2\,a_0$
vs. 2.14\,$a_0$ for CO and 2.45\,$a_0$ vs 2.28$\,a_0$ for OO). The
calculations place the COO minimum at 7.35\,eV, about
0.1\,eV below the lowest dissociation threshold.

The ground electronic state correlates adiabatically with
two dissociation channels,
\begin{eqnarray}
{\rm CO}_2 + h\omega (E_{\rm ph.} \ge \mbox{ }7.41\,{\rm eV}) & \rightarrow & 
{\rm O}(^1\!D) + {\rm CO}(X^1\Sigma^+)  \label{diss-chan1}\\
{\rm CO}_2 + h\omega (E_{\rm ph.} \ge \mbox{ }11.52\,{\rm eV}) & \rightarrow & 
{\rm C}(^3\!P) + {\rm O}_2(X^3\Sigma^-) \, ,  \label{diss-chan3}
\end{eqnarray}
and the ZPE corrected dissociation energies $D_0$ are shown in 
Table\ \ref{table1}. In channel (\ref{diss-chan1}), 
the calculated $D_0$ is 0.13\,eV less than the experimental value. 
The deviation might reflect a
large basis set superposition error introduced by the
diffuse functions and as such is the downside of the highly accurate 
vibrational spectrum in Table \ref{table2}. The error is independent of the
arrangement channel, and $D_0$  
in channel (\ref{diss-chan3}) [closed between 120\,nm and 160\,nm]
is equally underestimated. The calculations of Hwang and
Mebel\cite{HM00A}, using a noticeably smaller basis set, perfectly
agree with the experimental dissociation energy for this channel. 

One-dimensional (1D) cuts through the ground state PES 
are given for several $\alpha_{\rm OCO}$ angles 
in panels (a,c,e) of Figs.\ \ref{curco1} and \ref{curco2}. 
Black solid circles are the raw
adiabatic energies. The O + CO  limit is reached
smoothly and no barrier is detected towards the 
asymptote for any orientation of the CO diatom.
The same is true for the C + O$_2$ channel,
as illustrated in Fig.\ \ref{curco2}(e); 
the potential well in Fig.\ \ref{curco2}(e) is the COO isomer.
Angular dependence of the $\tilde{X}^1\!A'$ state is shown
in panels (a,c,e) of Figs.\ \ref{curwink1} and \ref{curwink2} 
for two sets of fixed CO bonds. In Fig.\ \ref{curwink1}, $R_1$ is
fixed at the FC value; in Fig.\ \ref{curwink2}, it is fixed 
close to the equilibrium of the bent OCO. Consequently, 
although the carbene-type minimum is perceptible in
all panels, it is best seen in   Fig.\
\ref{curwink2}.  As CO$_2$ bends, the adiabatic 
$\tilde{X}^1\!A'$  
state (black dots) forms a sharp narrowly
avoided crossing  with the state $2^1\!A'$ around 
$\alpha_{\rm OCO}=100^\circ$ (see, for example, Fig.\ \ref{curwink1}). 
A
dynamically meaningful representation for such nearly degenerate pairs
is diabatic rather than adiabatic. In fact, the black line in 
Figs.\ \ref{curco1} --- \ref{curwink2} depicts the 
$\tilde{X}^1\!A'$ state locally diabatized at bent
geometries as described in Sect.\ \ref{qd1}. This is the reason why the
black dots sometimes switch away from the black line and why lines of different
colors cross. In this locally
diabatic picture, the bent OCO minimum correlates with the 
state $2^1\!A'$ [purple line]. The transition state
separating the bent and the linear minima,  analyzed by 
Xantheas and Ruedenberg\cite{XR94} and Hwang and Mebel\cite{HM00A}
is thus the signature of this two-state intersection. 
The two-dimensional (2D) contour
plots of the PES of the locally diabatic $\tilde{X}^1\!A'$ state,
used in the above calculations of vibrational states, are shown in Fig.\
\ref{2dx}.

\subsection{Overview of the excited electronic states}\label{ex1}

Potentials of the excited electronic states are shown in Figs.\ \ref{curco1}
and \ref{curco2} along one CO bond for $A'$ and $A''$ 
symmetries. As with the $\tilde{X}$ state, the solid circles indicate
ab initio adiabatic energies. 
In the CO + O arrangement channel, which is the focus of the present 
investigation, all calculated states but one converge to the
dissociation threshold (\ref{diss-chan1}). The state $2^1\!A''$, 
correlating with $1^1\Sigma_u^-$ at the FC point, reaches the higher 
lying threshold
\begin{equation}
\label{diss-chan2}
{\rm CO}_2 + h\omega (E_{\rm ph.} \ge \mbox{ }11.46\,{\rm eV}) \rightarrow 
{\rm O}(^3\!P) + {\rm CO}(a^3\Pi) \, .
\end{equation}
In the C + O$_2$ arrangement channel [Fig.\ \ref{curco2}(e,f)], 
three electronic states, the $\tilde{X}$ 
state and the two components of the $\Pi$ state, correlate with channel
(\ref{diss-chan3}). Three other states ($\Sigma^-$ and $\Delta$) converge 
to the electronically excited fragments:
\begin{equation}
\label{diss-chan4}
{\rm CO}_2 + h\omega (E_{\rm ph.} \ge \mbox{ }13.75\,{\rm eV})  \rightarrow 
{\rm C}(^1\!D) + {\rm O}_2(^1\Delta) \, .
\end{equation}

Topographic hallmarks of the excited states can be exemplified using the
states $2,3^1\!A'$. In the FC region near linearity [Fig.\ \ref{curco1}(a)],
the states $2^1\!A'$ and $3^1\!A'$ form two
sharp avoided crossings near $R_1 = 2.2\,a_0$ and $R_1 = 2.8\,a_0$. These 
crossings are in fact two CIs between states 
$1^1\Pi_g$ and $1^1\Delta_u$.\cite{KRW88,SFCCRWB92,G12A,GB12} 
 The CIs 
are not independent: They are connected into a whole line,   
called a \lq CI seam',\cite{Y04} 
unusual properties of which\cite{GB12} are discussed in Sect.\
\ref{ex2}. As the molecule bends, the gap between the adiabatic states
grows. In the lower state $2^1\!A'$, the intersection cone first turns into
a broad barrier along the dissociation path 
[$\alpha_{\rm OCO} > 170^\circ$, Fig.\ \ref{curco1}(c)].
 As $\alpha_{\rm OCO}$ decreases further, the local minimum
near 2.3\,$a_0$ deepens and the dissociation barrier disappears
[Fig.\ \ref{curco1}(e) and Fig.\ \ref{curco2}(a)]. In
linear COO  [$\alpha_{\rm OCO}=0^\circ$, Fig.\ \ref{curco2}(e)], 
the sharp avoided crossing between 
states $2,3^1\!A'$ reappears again, this time at $R_1 = 3.5\,a_0$. 
Similar to $\tilde{X}$, the state $2^1\!A'$ supports a COO
intermediate, although the local minimum lies 0.7\,eV above the 
C + O$_2$ threshold (\ref{diss-chan3}). 
The evolution of the uppermost state $3^1\!A'$ with decreasing angle
is different, because its topography is very much influenced by a
pronounced barrier located outside the FC zone near 
$R_1 = 3.6\,a_0$ and separating the flat inner region
from a steep decline towards the asymptotic limit. This barrier is
distinct over a broad angular range and its sharpness
suggests a CI with a  higher lying state.
The nature of this intersection becomes 
apparent in Sect.\ \ref{ex5} discussing valence/Rydberg crossings. 

The $A''$ states are similar in many
respects. Near linearity, both symmetries
mirror the topography of the orbitally degenerate states
$1^1\Pi_g$ and $1^1\Delta_u$, the  
states $1,3^1\!A''$  are involved in the same CIs in the FC region  
[Fig.\ \ref{curco1}(b)], and their behavior in the bent molecule
up to $\alpha_{\rm OCO} = 120^\circ$ closely follows that of
$2,3^1\!A'$. The $1^1\!A''$ state stabilizes with
decreasing $\alpha_{\rm OCO}$, while the $3^1\!A''$ state
features a pronounced barrier around $R_1 = 3.6\,a_0$
[Fig.\ \ref{curco1}(d,f) and Fig.\ \ref{curco2}(b)]. Similarities persist
to $\alpha_{\rm OCO}=0^\circ$ [Fig.\ \ref{curco2}(f)]: 
The states $1,3^1\!A''$  form a CI at  $R_1 = 3.5\,a_0$, and the
state $1^1\!A''$ supports a local COO minimum. 
Differences with the $A'$ symmetry are
due to the state $2^1\!A''$, which in the FC region correlates with 
$1^1\Sigma_u^-$ and which has no counterpart among the calculated $A'$
states. This state, which at linearity is accidentally
degenerate with $1^1\Delta_u$, has a single minimum near
$2.4\,a_0$ and approaches the dissociation limit (\ref{diss-chan2})
without a barrier. This simple shape is preserved through most cuts
in Figs.\ \ref{curco1} and 
\ref{curco2}. The state $2^1\!A''$ also supports a COO minimum
[Fig.\ \ref{curco2}(f)]. 


Potential cuts along bending angle are shown in 
Figs.\ \ref{curwink1} and \ref{curwink2}. The doubly
degenerate $1^1\Pi_g$ and $1^1\Delta_u$ states
split into $A'$ and $A''$ components for $\alpha_{\rm OCO} < 180^\circ$. 
Evolution of the electronic energies with
decreasing $\alpha_{\rm OCO}$ was analyzed by Spielfiedel et al. using Walsh 
rules.\cite{SFCCRWB92}  Based on this analysis, 
the adiabatic excited states are commonly classified as bent or
linear. The energy of states $2^1A'$ and $1^1A''$
lowers as the molecule bends, while the energy of states  $3^1A'$ and
$3^1A''$ grows 
[see Fig.\ \ref{curwink1}(a,b)]. The global minima of the 
\lq bent' states lie near 120$^\circ$ ($2^1A'$) and 130$^\circ$
($1^1A''$) and are located below the dissociation limit (\ref{diss-chan1})
[Fig.\ \ref{curwink1}(a,b) and Fig.\ \ref{curwink2}(a,b)]. 
Bent equilibrium of the state 
$2^1A'$, which at $C_{2v}$ geometries becomes
$^1\!B_2$, was predicted by Dixon in his 
analysis of the CO flame emission bands.\cite{D63}  
Finally, the state $2^1A''$, which together with $3^1A'$ and
$3^1A''$ is \lq linear', 
is the least anisotropic of all states 
in a broad vicinity of $\alpha_{\rm OCO} \sim 180^\circ$. 

Properties of various stationary points in the PESs of
excited electronic states are given in
Table\ \ref{table3}. The $C_s$ point group notation is
used to label adiabatic states;  
$D_{\infty h}$ labels refer to the diabatic states at the FC point.
Experimental reference data for the excited electronic states are
scarce and fit into the footnote $a$.  For the bent 
states, the experimental\cite{D63,CLP92} 
equilibrium CO distances and the bending angle are 
reproduced within $0.1\,a_0$ and $3^\circ$, respectively; the bending
frequencies are accurate to within 30\,cm$^{-1}$ or better, 
and the calculated band
origins lie within the experimental uncertainties. 
Results of the previous ab initio studies\cite{KRW88,SFCCRWB92} are also
shown in Table\ \ref{table3}. Agreement in the equilibrium geometries
is excellent, with the exception of the
intersection-ridden uppermost states $3^1A'$ and $3^1A''$,
for which $C_{2v}$-restricted calculations of Refs.\
\onlinecite{KRW88} and \onlinecite{SFCCRWB92} miss the local $C_s$ minima
in the upper CI cones.  Vertical excitation energies $T_v$ are 
close to those of Ref.\ \onlinecite{SFCCRWB92}. Slight differences are  not
surprising for near degenerate states whose ordering  
is sensitive to the details of the 
ab initio set up. Peculiarity of the spectral region
120\,nm --- 160\,nm is that $T_v$ values 
are poor approximations to the positions of the absorption maxima 
because all transitions are electronically forbidden and the 
TDMs at the FC point are strictly zero. Finally, 
the ab initio dissociation energies $D_0$ in channels 
(\ref{diss-chan1}), (\ref{diss-chan3}), (\ref{diss-chan2}), and 
(\ref{diss-chan4}), 
also given in Table\ \ref{table3}, agree with the experimental
values\cite{HERZBERG67,AMS79} 
within $\sim 0.15$\,eV irrespective of the particular arrangement
or electronic channel.

\subsection{Topography of state intersections in the FC zone}\label{ex2}

Intersections of electronic states in linear CO$_2$ follow simple \lq
symmetry rules' \cite{KDC84,CMD84}  which severely restrict
the intersection topography.
Two types of intersections with regard to their symmetry properties
are prominent in Figs.\ \ref{curco1}(a,b) and \ref{curwink1}(a,b):\newline
\noindent (1) Renner-Teller (RT) glancing intersections\cite{R34,HERZBERG67} 
involve the $A'$ and $A''$ components of the $1^1\Pi_g$ or
the $1^1\Delta_u$ state.  In the rotating molecule, the
$A'/A''$ interaction  $\sim \Lambda\Omega/\sin^2\alpha_{\rm OCO}$
is proportional to the projections $\Lambda$ and $\Omega$ of the electronic 
($\hat{L}$) and  the total angular momentum 
($\hat{J} = \hat{N}_{\rm CO2} + \hat{L}$) 
on the molecular axis and diverges for 
$\alpha_{\rm OCO}\rightarrow 180^\circ$.\cite{R34,HERZBERG67,P88,GGH93}  
The $1^1\Pi_g \leftarrow \tilde{X}^1\Sigma_g^+$ transition 
is best classified as linear-linear, and
the $1^1\Delta_u \leftarrow \tilde{X}^1\Sigma_g^+$ transition 
is linear-bent. The \lq degeneracy manifold' for the
intersecting states is
the whole $(R_1,R_2)$ plane defined by the condition 
$\alpha_{\rm  OCO}= 180^\circ$.

\noindent (2) Two nested CIs involve the
$2,3^1\!A'$  and $1,2^1\!A''$  or
$2,3^1\!A''$ states and stem from the accidental $^1\Pi_g/^1\Delta_u$ crossing.
Both  degeneracies are 
lifted linearly along the tuning and coupling modes spanning 
the common branching space.\cite{Y04}  
According to Fig.\ \ref{curco1}(a,b), the tuning mode
is $R_{\rm CO}$, i.e. a combination of the  
symmetric and antisymmetric stretch (their irreps 
are $\sigma_{g,u}^+$ in the $D_{\infty h}$ group); $\alpha_{\rm OCO}$
breaking the linear symmetry (irrep $\pi_u$) is
the coupling mode. As a consequence of the  
\lq symmetry rules', the CIs occur along a line 
$F_{\rm CI}(R_1^\star,R_2^\star) = 0$ in the $(R_1,R_2)$ plane
at $\alpha_{\rm  OCO}= 180^\circ$: 
The \lq degeneracy manifold' is a 1D seam.\cite{HERZBERG67,T37,Y04}

The CI seam, constructed separately for the $2,3^1\!A'$ and $1,3^1\!A''$ states 
on a fine $(R_1,R_2)$ grid,\cite{GB12} 
is depicted in Fig.\ \ref{r1r2A}. It has two
remarkable properties. First, the intersection along the seam is
fivefold. Two CIs and two RT intersections imply four degenerate states. 
The hitherto ignored state $1^1\Sigma_u^-$ closely follows
$1^1\Delta_u$: The $\Delta/\Sigma$ energy gap
falls consistently below 300\,cm$^{-1}$ which is at the limit of the ab initio 
accuracy. Thus, the
$1^1\Pi_g/1^1\Delta_u$ and the  $1^1\Delta_u/1^1\Sigma_u^-$ pairs 
cross at the same CO bond distances and the total degeneracy is five. 
Second, the calculated seam traces out a closed loop. Closed CI seams
are rarely encountered,\cite{ARN97,GB12} although
arguments have been devised to prove their
ubiquity.\cite{LHVBZKB08,LHVBZKB09} A technical implication
of closed or strongly curved seams
is that local diabatization schemes fail\cite{GB12} and global 
or semiglobal diabatization\cite{KGM01} becomes necessary. The impact 
of closed seams on photodissociation dynamics has never been systematically
investigated.\cite{GB12} In CO$_2$, 
only a small portion of the loop is directly accessible to UV light. 
Nevertheless, paper II demonstrates that the seam topology deeply
affects the observed absorption spectrum. 

The curved intersection seam is responsible for a peculiar shape
of the potentials in the FC region. The adiabatic PESs of  the states
$2,3^1\!A'$ in linear CO$_2$ are shown in Fig.\
\ref{r1r2A}. The lower adiabatic state $2^1\!A'$ has a $C_{\rm 2v}$ 
minimum at $R_1 = R_2 = 2.41\,a_0$, and two $C_{\rm s}$ symmetric
saddles near $R_1 \approx 2.8\,a_0$ or $R_2 \approx 2.8\,a_0$, 
separating the  $C_{\rm 2v}$ minimum from the dissociation
asymptotics.  Outside the area enclosed by the seam line, 
the $2^1\!A'$ state has the $\Pi$, inside the $\Delta$ character. 
The saddles hide portions of cusp lines, along which the states
intersect and which are washed out by the
spline interpolation. The topography of the upper adiabatic $3^1\!A'$
state is a literal mirror image of the lower state; one finds 
a $C_{\rm 2v}$ saddle at $R_1 = R_2 = 2.41\,a_0$
and two $C_{\rm s}$ symmetric minima near 
$(R_1,R_2) = (2.84\,a_0,2.27\,a_0)$ and 
$(2.27\,a_0,2.84\,a_0)$. The state character changes from $\Delta$
outside the seam line to $\Pi$ inside. The two minima are
cusp-like and correspond to the upper cones of the CIs. 
The two other saddle points in the $3^1\!A'$ state, lying outside the
FC zone near $(R_1,R_2) = (3.6\,a_0,2.2\,a_0)$ and $(2.2\,a_0,3.6\,a_0)$, 
result from an avoided crossing with a higher
lying state. 

Changes of the electronic character of $A''$ states across the
intersection can be monitored using matrix elements of the electronic angular
momentum $\hat{L}_z$.\cite{DK97} 
The matrix elements $\left|\langle iA''|\hat{L}_z|jA'\rangle\right|$ along 
the line passing twice through the closed seam 
are depicted in Fig.\ \ref{mrcilz} for $i = 1, 2, 3$ and $j = 2,3$. 
The states $|2^1\!A'\rangle$ and $|3^1\!A'\rangle$ act as \lq probes'
whose assignment in terms of $\Pi$ or $\Delta$  is known. Outside the
seam area, most matrix elements are close to integer values of 0, 1, and 2. 
For $R_1 \le 2.2\,a_0$ for example, 
$\left|\langle 1A''|\hat{L}_z|2A'\rangle\right| = 1$ and 
$\left|\langle 2A''|\hat{L}_z|2A'\rangle\right| = 0$ 
so that $|1A''\rangle$ is a $\Pi$ state, while 
$|2A''\rangle$ is a $\Sigma$ state.
$\left|\langle 3A''|\hat{L}_z|2A'\rangle\right|$ vanishes too, but the
state $|3A''\rangle$ is a $\Delta$ state, as confirmed by
the matrix element $\left|\langle 3A''|\hat{L}_z|3A'\rangle\right| =
2$. As the CI seam is crossed at $R_1 = 2.25\,a_0$ into the area interior
to the seam loop,  the three-state intersection induces violent 
changes in the electronic labels of the adiabatic
states. The state $|3A''\rangle$ becomes a $\Pi$ state  
($\left|\langle 3A''|\hat{L}_z|3A'\rangle\right| = 1$); the state 
$|2A''\rangle$ acquires $\Delta$ character 
($\left|\langle 2A''|\hat{L}_z|2A'\rangle\right| \approx 2$), while the
state $|1A''\rangle$ acquires $\Sigma$ character 
($\left|\langle 1A''|\hat{L}_z|2A'\rangle\right| \approx 0.5$), and the
non-integer values of the latter two reflect strong
mixing of the near degenerate $\Delta_u/\Sigma^-_u$ pair. 
The next reshuffling occurs as the seam loop is crossed outwards at
$R_1 \approx 2.8\,a_0$. The state $|1A''\rangle$ becomes $\Pi$ again, 
while the matrix elements involving $|2A''\rangle$ and $|3A''\rangle$
vary until $R_1 \approx 3.6\,a_0$ is reached, where
$\left|\langle 3A''|\hat{L}_z|3A'\rangle\right|$ finally vanishes
indicating that $|3A''\rangle$ emerges from the crossing
region as a $\Sigma$ state converging towards the upper threshold 
(\ref{diss-chan2}), while $|2A''\rangle$ becomes $\Delta$ state
correlating with threshold (\ref{diss-chan1}). 

The positions of cusps, minima, and saddles in the PESs of the 
$A''$ states at linearity are identical to those in $A'$
states. Exception is
the PES of the state $2^1\!A''$, correlating in the FC region with
$1^1\Sigma_u^-$. Its appearance, illustrated in Fig.\ \ref{2d2app},  
is very much simplified by it being tightly linked with the diabatic 
$1^1\Delta_u$ state. The PES 
has a single $C_{\rm 2v}$ minimum at $R_1 = R_2 = 2.41\,a_0$, 
is free from cusps seen in other adiabatic PESs, and
provides a nature's illustration of the
shape of the diabatic $1^1\Delta_u$ PES. 

CIs at linearity can also be recognized in the potentials plotted in
the $(R_1,\alpha_{\rm OCO})$ plane. 
An example including four electronic states at  $R_2 = 2.2\,a_0$ is 
shown in Fig.\ \ref{r1gaA}. Although the characteristic features are
no longer $C_{2v}$ symmetric, a maximum in the  $2^1\!A'$ state
at $\alpha_{\rm OCO} = 180^\circ$ and $R_1 = 2.8\,a_0$ 
and two minima in the  $3^1\!A'$ state
at $\alpha_{\rm OCO} = 180^\circ$ and $R_1 = 2.3\,a_0$  and 
$R_1 = 2.8\,a_0$ are recognizable in panels (a) and (b). Close
to linearity, the $A''$ states [panels (c) and (d)] maintain the same
contour maps as $A'$ states. 

\subsection{Intersections and avoided crossings in bent CO$_2$}
\label{ex4}

As CO$_2$ bends and all above degeneracies are removed, 
the primary topographic features (cusps, minima, and saddles) 
remain clearly visible up to 
$\alpha_{\rm OCO} \sim 160^\circ$. This is demonstrated in the
$(R_1,\alpha_{\rm OCO})$ plane in Fig.\ \ref{r1gaA} and in the
$(R_1,R_2)$ plane  in Fig.\ \ref{r1r2B}. Further decrease of
$\alpha_{\rm OCO}$ spawns new avoided crossings, a detailed map of
which is given in the angular cuts in Figs.\ \ref{curwink1} and
\ref{curwink2}. All pairs of
calculated states become near degenerate at various geometries. For example, 
the $\tilde{X}$ state forms avoided crossings successively with states
$2^1\!A'$ at $\alpha_{\rm OCO} = 100^\circ$ and $3^1\!A'$ at 
$\alpha_{\rm OCO} = 80^\circ$ (Fig.\ \ref{curwink2}). The state
$3^1\!A''$ approaches closely $2^1\!A''$ at 
$\alpha_{\rm OCO} \sim 120^\circ-130^\circ$ and $1^1\!A''$ at 
$\alpha_{\rm OCO} = 90^\circ$, and the states  $1^1\!A''$ and $2^1\!A''$ 
cross between $70^\circ$ and $80^\circ$.  Solid lines in  Figs.\ \ref{curwink1}
and \ref{curwink2} cross because they refer to the states 
locally diabatized through the \lq bent' avoided crossings (see Sect.\
\ref{qd1}).  Potential curves of the states $3^1\!A'$ and
$3^1\!A''$ bear evidence of strong interactions with  the next higher
states. Especially pronounced in 
Fig.\ \ref{curwink1}(d,f) and Fig.\ \ref{curwink1}(b,d) 
are the sharp near-intersections in $3^1\!A''$ around 155$^\circ$.

As a result of multiple avoided crossings,  
local bent equilibria are found in all calculated
states. The diabatic origin of a particular local minimum is
invariably different from the adiabatic one.
Consider the carbene-like bent OCO with  
$\alpha_{\rm OCO} \sim 70^\circ$ 
in Fig.\ \ref{curwink2}(a). Purely adiabatically
(solid dots), this minimum belongs to the ground electronic state. In
the locally diabatic picture, bent OCO belongs the state 
$2^1\!A'$ (purple line). An avoided crossing 
between states $2^1\!A'$  and $3^1\!A'$, recognizable in 
panel (a) around $\alpha_{\rm OCO} \sim 95^\circ$, implies that another
diabatization might  re-assign the bent OCO minimum
to the $3^1\!A'$ state (brown line). Furthermore, the
broad barrier in $3^1\!A'$ near $\alpha_{\rm OCO} \approx 140^\circ$
indicates another avoided crossing with the next higher $A'$ state which thus 
 would be the true owner of the bent OCO minimum. Similar analysis
 applies to the bent OCO in the $A''$ states in
Fig.\ \ref{curwink2}(b). Depending on the chosen representation, 
it can be ascribed to the fully  adiabatic state $1^1\!A''$ (dots), to
the locally diabatic state $3^1\!A'$ (brown line), or --- via the sharp
near-intersection around $155^\circ$ --- to the
next higher $A''$ state. Other bent conformations in
the excited states result from state interactions, too. One such isomer with the
valence angle of 100$^\circ$ is created via
an avoided crossing in $3^1\!A'$ [Fig.\ \ref{curwink1}(c,e)].
Crossing of the same $3^1\!A'$ state with the next higher state
between 60$^\circ$ and 70$^\circ$ leads to another high-energy bent minimum 
[Fig.\ \ref{curwink2}(a,e)].

\subsection{Beyond the first six states:  
Valence/Rydberg crossings}\label{ex5}

Many local barriers and minima in the states 
$3^1\!A'$ and $3^1\!A''$ 
are due to intersections with \lq invisible' higher lying electronic states. 
Previous detailed ab initio studies exposed these 
\lq invisible' intersection partners as mainly Rydberg 
states.\cite{EEW77,CJL87,SFCRW91,SFCRW93} 
The aim of this section is to illustrate how the Rydberg/valence 
interaction affects the potential profiles  along the dissociation
path and to outline the geometries at which valence states can
effectively drain population from the Rydberg manifold. 
To this end, CASSCF calculations of the first 10 states, 
$1$---$5^1\!A'$ and $1$---$5^1\!A''$,  have been performed with 
the d-aug-cc-pVQZ basis set for the CO bond
distances $R_{\rm 2} \in [1.7\,a_0,5.0\,a_0]$ and the OCO bond angles 
$\alpha_{\rm OCO} \in [100^\circ,179^\circ]$. One CO bond is
kept fixed at $R_1 = 2.4\,a_0$, so that only CO + O
arrangement channel is covered.

The CASSCF potentials along $R_2$ 
are shown in Fig.\ \ref{mcscfr1}. All five $A''$ states and
four $A'$ states converge to thresholds 
(\ref{diss-chan1}) or (\ref{diss-chan2}). 
The states correlating with the lowest threshold (\ref{diss-chan1}) are of 
$\Sigma^+$, $\Delta$, and $\Pi$ symmetry. The states
correlating with the highest threshold  (\ref{diss-chan2}) are
$\Sigma^\pm$, $\Delta$, and $\Pi$. One $A'$ state converges towards
the asymptote
\begin{eqnarray}
{\rm CO}_2 + h\omega (E_{\rm ph.} \ge \mbox{ }9.64\,{\rm eV}) & \rightarrow & 
{\rm O}(^1\!S) + {\rm CO}(X^1\Sigma^+) \, . \label{diss-chan5}
\end{eqnarray}
At linearity, this state is the Rydberg
state $1^1\Sigma^+_u(3\pi_u)$.\cite{NOTE-CO23-1} 
Compared to Fig.\ \ref{curco1}, many new avoided crossings 
are found in Fig.\ \ref{mcscfr1}(a,b). 
The $D_{\infty h}$ notation of a state $|i\rangle$ 
in these panels is related to the
expectation value $\langle i| L_z^2 | i \rangle$ which a 
diabatic state preserves across the intersections. 
Potential curves in Fig.\ \ref{mcscfr1}(a,b)
are color coded according to the $\langle L_z^2 \rangle$ values:   
$\Sigma$ states ($\langle L_z^2\rangle = 0$) are shown in green; 
$\Pi$ states ($\langle L_z^2\rangle = 1$) are blue, and 
$\Delta$ states ($\langle L_z^2\rangle = 4$) are red/orange. 
Clearly, most adiabatic curves change
color more than once as the CO bond stretches: 
They are tailored out of several diabatic states.

Familiar from the discussion in Sect.\ \ref{ex2} are the fivefold
crossings involving $1^1\Pi_g(5\sigma_g)$ and the accidentally
degenerate pair $1^1\Delta_u(2\pi_u)/1^1\Sigma_u^-(2\pi_u)$. These
crossings are marked with arrows in Fig.\ \ref{mcscfr1}(b). As in the
MRD-CI calculations, the $\Delta_u/\Sigma_u^-$ pair is easily
recognizable using matrix elements of $L_z$. At linearity, $\langle
L_z^2\rangle$  for these two states is either 0 or 4, but already a tiny
deviation of 2$^\circ$ 
causes $\langle L_z^2\rangle$ to collapse towards an average
value of $\sim 2$. This state mixing is stressed in Fig.\ \ref{mcscfr1}(a,b)
with the $\Delta_u/\Sigma_u^-$ label and with the red/green color. 
The second edition of the mixed  $\Delta_u/\Sigma_u^-$ 
pair is found around 11.5\,eV. 
The Rydberg state $2^1\Delta_u(3\pi_u)$, shown with an orange
line in Fig.\ \ref{mcscfr1}(a,b), carries as a satellite the state 
$2^1\Sigma_u^-(3\pi_u)$ [green line in Fig.\ \ref{mcscfr1}(b)]. Again,  
$\langle L_z^2\rangle$ for these states assumes a  
non-integer value between 1.5 and 2.5 in even slightly bent molecule. 

The valence pair $1\Delta_u/1\Sigma_u^-$  converges towards the 
uppermost threshold (\ref{diss-chan2}) and successively traverses
the higher lying states. The prominent dissociation 
barrier in the $3^1\!A'$ and $3^1\!A''$ states near $R_2 = 3.6\,a_0$ 
is a clear signature of the sixfold valence/Rydberg crossing 
involving $1\Delta_u/1\Sigma_u^-$, the  
$1^1\Sigma_u^+$ state [green line in Fig.\ \ref{mcscfr1}(a)] 
and a repulsive $R^1\Delta_u$ state [orange line in Fig.\ \ref{mcscfr1}(a,b)]. 
$1^1\Sigma_u^+$ is the optically bright 
Rydberg state responsible for the strong absorption band around 
11.1\,eV (111.7\,nm).\cite{CJL87} The state $R^1\Delta_u$ corresponds to a pair
of strongly repulsive $^1\!A'/^1\!A''$ states descending
towards threshold (\ref{diss-chan1}) 
from very high energies. Two sections of this repulsive potential curve 
are seen in Fig.\ \ref{mcscfr1}(a,b) in the intervals  $[3.2\,a_0,3.7\,a_0]$ and
$[3.7\,a_0,5.0\,a_0]$. Thus, the asymptotic repulsive portions of the
$3^1\!A'$ and $3^1\!A''$ states diabatically belong to $R^1\Delta_u$. 
The sixfold crossing 
$1\Sigma_u^+/R^1\Delta_u/1\Delta_u/1\Sigma_u^-$ near $R_2 = 3.6\,a_0$ 
is not the only one involving $R^1\Delta_u$ state. At higher
energies and shorter CO bonds, $R^1\Delta_u$  intersects the Rydberg 
pair $2\Delta_u/2\Sigma_u^-$  (cf. Fig.\ \ref{mcscfr1}(a,b)
near $R_2 \approx 3.0\,a_0$). The diabatic $R^1\Delta_u$ state distinctly 
stands out because it strictly preserves the projection 
$\langle L_z^2\rangle \approx 4$ even for $\alpha_{\rm OCO} =
165^\circ$, while $\langle L_z^2\rangle$ values for the other states
become non-integer. 

The above discussion is valid for both $A'$ and $A''$ states. The
$A''$ symmetry block in Fig.\ \ref{mcscfr1}(b) contains one more
state, namely the Rydberg state $2^1\Pi_g(6\sigma_g)$ missing among the
$A'$ states where it would have been $6A'$. This state, materializing 
out of nowhere at $R_2 = 3.0\,a_0$, is involved in the 
$R^1\Delta_u/2\Delta_u/2\Sigma_u^-$ crossing --- making it a
sevenfold intersection.

While the adiabatic gaps in the valence/valence intersection
$1\Pi_g/1\Delta_u/1\Sigma_u^-$ grow as $\alpha_{\rm OCO}$ deviates
from 180$^\circ$, all Rydberg/Rydberg and Rydberg/valence
intersections not only remain recognizable in bent CO$_2$, but clearly
sharpen up. This is illustrated in panels
(c) --- (f) of Fig.\ \ref{mcscfr1} drawn for $\alpha_{\rm OCO} = 175^\circ$ and
160$^\circ$. As a side result, the repulsive $R^1\!A'$ and $R^1\!A''$
states, deriving from $R^1\Delta_u$, remain distinct at all 
angles, despite new \lq bent' intersections in the
$3,4,5^1A'$ and $3,4,5^1A''$ states [see, 
for example, Fig.\ \ref{mcscfr1}(e,f)].\cite{NOTE-CO23-2}
Via these intersections,
carbon dioxide excited with vacuum UV light can --- at bent
geometries --- reach any of the shown dissociation channels. 
Although a detailed analysis of these multiple 
pathways is beyond the scope of the present 
work, a brief description of intersections involving the
optically bright $^1\Sigma_u^+$ state is added at the end of this
section.

Potential curves along $\alpha_{\rm OCO}$ are shown in
Fig.\ \ref{mcscfw1}. The crossing
patterns, expecially numerous among $A''$ states, explain intricate 
topography of states $3^1\!A'$ and $3^1\!A''$ in Fig.\
\ref{curwink1}. For example, the sharp barrier in the state $3^1\!A''$ at
angles between 150$^\circ$ and 170$^\circ$ [Fig.\ \ref{curwink1}(d,f)
and Fig.\ \ref{curwink1}(b,d)] originates from a complicated
three state crossing around 165$^\circ$  involving states
$3,4,5^1\!A''$ [Fig.\ \ref{mcscfw1}(b)]. 
Diabatically, the decreasing branch of the
$3^1\!A''$ state at $\alpha_{\rm OCO} < 160^\circ$ belongs to
$4^1\!A''$ --- the state which at linearity merges into the 
Rydberg pair $2\Delta_u/2\Sigma_u^-$. This diabatic 
state can be followed to even smaller angles 
through another intersection, this time  
with $2^1\!A''$ near $130^\circ$. Ultimately, as indicated with an
arrow in Fig.\ \ref{mcscfw1}(b), it is this diabatic
state which the carbene-like OCO minimum belongs to.
The avoided
crossings in the $A'$ symmetry block are similar and simpler. The
three state crossing of $3,4,5^1\!A''$ is a mere ghost
because the adiabatic gap is almost 2\,eV wide and the barrier in the
$3^1\!A'$ state is broad and low. The state $4^1\!A'$, originating
from $2^1\!\Delta_u$, stabilizes upon bending and is
easily traced to smaller angles through the 
intersection with $3^1\!A'$ near $\alpha_{\rm OCO} =
150^\circ$. Again, the  carbene-like OCO correlates diabatically 
with the state originating from $2^1\!\Delta_u$ at linearity. 

The analysis of broadening and splittings
in the strong absorption band of the Rydberg state  $1^1\Sigma_u^+$ 
focused on the interactions of $1^1\Sigma_u^+$ 
with valence states.\cite{CJL87,SFCRW91,SFCRW93}
Present calculations reveal numerous
avoided crossings with both valence and Rydberg states inside and outside the
FC zone. At linearity, the $1^1\Sigma_u^+$ state cuts twice through the
Rydberg state $2^1\!\Delta_u$ [Fig.\ \ref{mcscfr1}(a)], and
both crossings persist to smaller angles 
[Fig.\ \ref{mcscfr1}(c,e)]. Another crossing occurs near $R_2
= 3.7\,a_0$ (for all calculated angles) 
and mixes $1^1\Sigma_u^+$ with $R^1\Delta_u$.  This
interaction strongly perturbs the ab initio 
$\langle L_z^2\rangle$ values but vanishes  beyond $R_1 = 4.0\,a_0$. 
The potential cut along $\alpha_{\rm OCO}$ in Fig.\
 \ref{mcscfw1}(a) exposes another avoided crossing at 
$\sim 120^\circ$ which involves the states $4,5^1\!A'$ and leads to a local
 minimum in $5^1\!A'$ (the state, correlating with
$1^1\Sigma_u^+$ at these R$_{\rm CO}$). Finally, the 
barrier in $5^1\!A'$ near $150^\circ$ implies interaction  
with the next higher state, which according to the analysis of
Ref.\ \onlinecite{SFCRW93} 
correlates with $2^1\Sigma_u^+$ at linearity. While all these
crossings can redirect population from the 
optically bright $1^1\Sigma_u^+$ state along various linear and bent 
routes, Fig. \ref{mcscfr1} clearly demonstrates that the diabatic
$1^1\Sigma_u^+$ state, calculated at the CASSCF level, is repulsive at all 
geometries and thus can dissociate directly.

\section{Quasi-diabatization of the valence states}
\label{diabats}

As explained in the Introduction, the diabatic 
representation,\cite{NOTE-CO21-3} although not generally indispensable, 
is best suited for the particular implementation of nuclear quantum
dynamics used in paper II. Rigorously speaking, 
diabatization has to be performed simultaneously on all six calculated valence
states, because each pair of states intersects either in
the FC zone or at bent geometries.
Simplifications to this \lq Herculean task'\cite{QZGSCH05}
stem from the expectation that the two groups  of intersections 
influence photodissociation in different ways: While the
\lq bent' CIs affect later stages of the product formation, 
the electronic branching ratios, and/or the rovibrational
photofragment distributions,\cite{CO2-5} 
the \lq FC' CIs are directly responsible for the 
shape of the observed absorption spectrum. This distinction guides the
practical construction of the diabatic states:
The \lq bent' CIs and the \lq FC' CIs are analyzed
at two different levels of detail, commensurate with their expected impact on
the absorption spectrum. As a result, the complete multistate 
problem splits into several steps and the need for a global diabatization of six
 multiply intersecting states is obviated.

\subsection{Intersections at bent geometries: Local diabatization}\label{qd1}

The vicinity of \lq bent' CIs is diabatized locally, using the
energy-based scheme as described, for example, by 
K\"oppel.\cite{K04} All considered CIs are shown in Figs.\
\ref{curwink1} and \ref{curwink2} and include 
\begin{enumerate}
\item[(a)] the
$\tilde{X}^1\!A'/2^1\!A'$ pair at $\alpha_{\rm OCO} = 90^\circ -
110^\circ$;
\item[(b)] the 
$2^1\!A''/3^1\!A''$ pair at $\alpha_{\rm OCO} =  120^\circ - 150^\circ$; 
\item[(c)] the $1^1\!A''/3^1\!A''$ pair, with  $3^1\!A''$ state diabatized in 
step (b), at $\alpha_{\rm OCO} =  70^\circ - 100^\circ$. 
\end{enumerate}
The $\tilde{X}^1\!A'/2^1\!A'$ pair is considered as an example. At
$C_{2v}$ geometries, these states belong to $A_1$ and $B_2$
irreps, so that their CI is \lq symmetry allowed'.\cite{Y04} Its
branching space
includes $\alpha_{\rm OCO}$ as the tuning mode (irrep $A_1$)
and the antisymmetric stretch $R_-$ as the symmetry breaking coupling  
mode (irrep $B_2$). 
The $2\times 2$ diabatic potential matrix ${\bf V}^d$ is constructed on the ab
initio grid of bond distances $(R_1,R_2)$ from
the diagonal adiabatic potential matrix ${\bf V}^a$ using the
orthogonal adiabatic-to-diabatic transformation (ADT)\cite{K04} 
\begin{equation}
\label{diaeq1}
{\bf V}^d = {\bf S}^T  {\bf V}^a {\bf S}\, ,
\end{equation}
with the transformation matrix  ${\bf S}$ 
parameterized by the mixing angle $\Theta$,
\begin{equation}
\label{diaeq2}
\Theta(R_1,R_2,\alpha_{\rm OCO}) = -\frac{1}{2}
\arctan{\left(\frac{\alpha_{\rm OCO}-
\alpha^\star_{\rm OCO}(R_1,R_2)}{W_0(R_1,R_2)}\right)} \, ,
\end{equation}
which varies between 0 and $\pi/2$; 
$\alpha^\star_{\rm OCO}(R_1,R_2)$ denotes the crossing point of the
diabatic potentials ($\Theta = \pi/4$); 
the width $W_0$ of the function $\Theta(\alpha_{\rm OCO})$ is
evaluated from the adiabatic energy gaps $\Delta V^a/2$ and the average slope
of the diabatic potentials $\overline{F}$, 
\begin{equation}
\label{diaeq3}
W_0 = \left. \frac{\Delta
    V^a}{2\overline{F}}\right|_{\alpha^\star_{\rm OCO}} \, ,
\end{equation}
at each grid point $(R_1,R_2)$ for $\alpha^\star_{\rm OCO} = 100^\circ$. 
With the off-diagonal coupling in Eq.\ (\ref{diaeq1}) neglected, 
the ground state $\tilde{X}^1\!A'$ becomes completely
decoupled from other $A'$ states. 
The remaining diagonal matrix element in ${\bf V}^d$ is the potential
waiting to be further diabatized at linearity.
The diabatization is kept local by restricting the
interval, over which $\Theta(\alpha_{\rm OCO})$ varies from 0 to
$\pi/2$, to the vicinity of $\alpha^\star_{\rm OCO}$: 
\begin{equation}
\label{diaeq3a}
\tilde{\Theta}(R_1,R_2,\alpha_{\rm OCO}) = g_1\left[g_2
\Theta(R_1,R_2,\alpha_{\rm OCO}) + \frac{\pi}{2}(1-g_2)\right] \, ,
\end{equation}
The two switching functions $g_{1,2}(\alpha_{\rm OCO})$,\cite{HH00B}
\begin{equation}
\label{diaeq4}
 g_{1,2}(\alpha_{\rm OCO}) = \left[1 + 
\exp(-(\alpha_{\rm OCO}-\alpha_{1,2})/\lambda_{1,2})\right]^{-1} \, ,
\end{equation}
with $\alpha_1 = 95^\circ$, $\alpha_2 = 105^\circ$, and 
$\lambda_1 = -\lambda_2 = 5^\circ$, are adjusted to restrict the
diabatization to angles 90$^\circ$---$110^\circ$. 
If $(\alpha_1-\alpha_2) \rightarrow 0$ in $g_{1,2}$, 
the so-called \lq diabatization by eye' is recovered, which
corresponds to a relabeling of ab initio energies at the crossing angle 
$\alpha_{\rm OCO}^\star$.\cite{LLYM94,GQZSCH07,G12A} 

The main advantage of local diabatization, the results of which
are shown in 1D in Figs.\ \ref{curwink1} and \ref{curwink2}, is that  
smooth potentials 
are created for strongly bent geometries at relatively low cost. The
quality of the resulting PESs
in 2D, either in the $(R_1,R_2)$ or
in the $(R_1,\alpha_{\rm OCO})$ plane is illustrated for all states 
in Figs.\ \ref{2dx}, \ref{2d2app}, \ref{r1gaA}, \ref{r1r2B},
and \ref{r1gaB}. For the 
lowest excited states in each symmetry block, $2^1\!A'$ and
$1^1\!A''$, the global bent equilibrium
at 120$^\circ$ or 130$^\circ$ [Fig.\ \ref{r1gaA}(a,c)] is not affected
by the local diabatization. In contrast, 
the carbene-like bent OCO minima change the owner: They appear in
the state  $2^1\!A'$ [local minimum near 70$^\circ$
barely visible in Fig.\ \ref{r1gaA}(a) but pronounced in  Fig.\
\ref{r1gaB}(a)] and in the state $3^1\!A''$ [Fig.\ \ref{r1gaB}(d)]. 
Barriers and local minima arising in the avoided crossings with
Rydberg states are not diabatized within this scheme. 
Examples are the sharp barrier in $3^1\!A''$ near 150$^\circ$ 
[Fig.\ \ref{r1gaA}(d) and Fig.\ \ref{r1gaB}(d)] and the deep local
minimum in $3^1\!A'$ near 100$^\circ$ [Fig.\ \ref{r1gaA}(b) and 
Fig.\ \ref{r1gaB}(b)]. 

\subsection{Intersections at linear geometries: Regularized 
diabatic states}\label{qd2}

In the second step, the CIs in the FC region are diabatized; 
kinematic singularities due to glancing intersections  
between $A'$ and $A''$ states are not considered.
Accurate diabatization schemes rely on direct numerical differentiation
of the electronic wave functions with respect to nuclear coordinates 
or on the orbital rotation method\cite{DW93,SHW99} 
as implemented in MOLPRO.\cite{MOLPRO-FULL} Both types of  calculations 
 become prohibitively expensive with the d-aug-cc-pVQZ basis set, and 
further approximations are needed in order to find the diabatic
potential matrix, consisting of a $2
\times 2$ block of  $A'$ and a $3\times 3$ block of $A''$ states:
\begin{equation}
\label{diarep1}
{\bf V}^{d} = \left(\begin{array}{cc|ccc} 
V_{\Delta'} & V_{\Pi'\Delta'}  & & & \\
V_{\Pi'\Delta'} & V_{\Pi'}& & & \\ \hline
 & & V_{\Sigma''}& V_{\Sigma''\Pi''} & V_{\Sigma''\Delta''}\\
 & & V_{\Sigma''\Pi''} & V_{\Pi''} 
& V_{\Pi''\Delta''} \\
 & & V_{\Sigma''\Delta''} & V_{\Pi''\Delta''} & V_{\Delta''}\\
\end{array}\right) \, .
\end{equation}
The five diagonal matrix elements are the five 
diabatic states $\Sigma''$, $\Pi'$, $\Pi''$, $\Delta'$, and $\Delta''$, 
with the smooth intersecting potentials in the $(R_1,R_2|\alpha_{\rm
  OCO} = 180^\circ)$ plane constructed to 
coincide at $\alpha_{\rm  OCO} = 180^\circ$
with the ab initio PESs for $1^1\Sigma^-$, $1^1\Pi_g$, and $1^1\Delta_u$,
respectively. 
The semi-global diabatization scheme, akin to the vibronic coupling 
model\cite{KDC84,K04} and adjusted to the topography of the closed fivefold CI
seam, has already been introduced in Ref.\
\onlinecite{GB12}. Diabatization proceeds in two steps. First,
a model diabatic matrix of the form Eq.\ (\ref{diarep1}) with 
elements $V_{ij}^{(0)}$ is constructed. 
Due to orbital degeneracy, $V^{(0)}_{\Pi'}(R_1,R_2) = V^{(0)}_{\Pi''}(R_1,R_2)$ and
$V^{(0)}_{\Delta'}(R_1,R_2) = V^{(0)}_{\Delta''}(R_1,R_2)$. Moreover, the
accidental near-degeneracy of states $1^1\Sigma_u^-$  and $1^1\Delta_u$ 
implies $V^{(0)}_{\Sigma''}(R_1,R_2) \approx V^{(0)}_{\Delta''}(R_1,R_2)$
in a broad vicinity of the closed CI seam --- the property
which substantially simplifies diabatization of $A''$ states.
Deviations from linearity, measured by the 
coordinate $Q_u \sim \sin\alpha_{\rm OCO}$, are included in the
model via off-diagonal matrix elements represented as symmetry adapted 
expansions in $Q_u$:
\begin{eqnarray}
\label{diarep2}
V^{(0)}_{\Pi'\Delta'} &  = & \sum_{k=0}^{N'} \alpha_k(R_1,R_2) Q_u^{2k+1}\\
V^{(0)}_{\Sigma''\Pi''} &  = &  V^{(0)}_{\Pi''\Delta''} = 
\sum_{k=0}^{N''} \beta_k(R_1,R_2) Q_u^{2k+1}  \, .
\end{eqnarray}
Couplings of the accidentally degenerate $\Sigma''$  and $\Delta''$ states to
$\Pi''$ are set equal, while the matrix element
$V^{(0)}_{\Sigma''\Delta''}$ for the RT-like $\Sigma''/\Delta''$ interaction
is neglected. The model is complete after the 
expansion coefficients in  $V_{ij}^{(0)}$ are calculated
from a non-linear least-squares fit to
ab initio energies.  In the second step, the regularized diabatic states 
approach is invoked,\cite{K04} and the  matrix elements $V_{ij}^{(0)}$ 
are used to define the orthogonal ADT matrix, which is applied to the 
adiabatic matrix ${\bf V}^a$ via Eq.\ (\ref{diaeq1}) giving the
desired diabatic matrix elements of Eq.\ (\ref{diarep1}) on the full
ab initio grid. Final interpolation is performed using 3D splines. 

The diabatization is localized to the vicinity of the
CI seam by modifying
matrix elements  $V^{(0)}_{ij}$ ($i\ne j$) in Eq.\ (\ref{diarep2}), 
\begin{equation}
\label{diarep3}
\tilde{V}^{(0)}_{ij} = 
\frac{V^{(0)}_{ij}}{1+\exp{\left[(\rho-\rho_0)/\lambda_0\right]}} \, ,
\end{equation}
with $\rho = \left[(R_1-2.605\,a_0)^2 + (R_2-2.605\,a_0)^2\right]^{\frac{1}{2}}$
being a radial distance from the center of a circle enclosing
the seam; $\rho_0 = 1.0\,a_0$ and $\lambda_0 = 0.1$. 
Adiabatic and diabatic states are forced to 
coincide if either bond becomes longer than $\sim 3.8\,a_0$, so that 
diabatization regions for linear and bent CIs are cleanly kept apart. 

The constructed representation describes best 
the vicinity of linearity in which non-adiabatic
transitions occur. The exact range of validity is determined by
the length of the expansion in Eq.\ (\ref{diarep2}). 
The choice $N' = N'' = 1$ gives a model which 
fits ab initio data with a root mean square error of $\sim 180$\,cm$^{-1}$
for angles $\alpha_{\rm OCO} \ge 150^\circ$.  
The ADT, constructed using this model, is guaranteed to remove kinematic
singularities at the CIs, but leaves the strength
of residual non-adiabatic couplings unspecified.\cite{KGM01}  
The ultimate test of the scheme is the quantum mechanical 
absorption spectrum described in paper II. In order to assess the
accuracy of the truncated $Q_u$ expansion,  three diabatic representations
are constructed, based on the expansion coefficients obtained from fitting
in three different angular ranges 
$180^\circ-170^\circ$, $180^\circ-160^\circ$, and $180^\circ-150^\circ$. 
The corresponding absorption spectra are virtually identical. 
The spectra are also insensitive to small variations in $\rho_0$ and
$\lambda_0$ in Eq.\ (\ref{diarep3}) --- the modifications take place 
too far away from the crossing seam to affect nuclear dynamics. 

Another test of the constructed ADT is given in 
Fig.\ \ref{theta}, in which the model mixing
angle $\tilde{\Theta}$ 
for states $2,3^1\!A'$ is compared with the ab initio one
calculated using MOLPRO with a smaller  cc-pVQZ atomic basis set.  
The dependence $\tilde{\Theta}(R_1)$ on the CO bond length 
has a characteristic bell shape: 
The closed CI seam is intersected twice giving rise to the ascending and
the descending branch. The curve $\tilde{\Theta}(R_1)$ 
flattens out as $\alpha_{\rm
  OCO}$ decreases and CO$_2$ leaves the degeneracy plane at
180$^\circ$. Agreement between the model
and the ab initio results is satisfying for all angles. A
constant shift of  25$^\circ$ applied to the ab initio mixing angle 
has no effect on the strength of non-adiabatic coupling proportional to
$\partial\tilde{\Theta}/\partial R_1$. 

One-dimensional cuts through the diabatic PESs (diagonal
matrix elements corresponding to states $1^1\Pi_g$ and $1^1\Delta_u$) 
are shown in Fig.\ \ref{1ddia}. They  
 cross at all geometries and can be directly compared
with adiabatic states in Fig.\ \ref{curco1}. The off-diagonal
coupling matrix elements are large in the intersection region
vanishing off towards the asymptotic channels. 
Diabatic matrix elements are
further illustrated in Fig.\ \ref{2ddia1} in
the $(R_1,R_2)$ plane and in Fig.\ \ref{2ddia2} in the $(R_1,\alpha_{\rm OCO})$
plane. In all cuts, the diabatic PESs smoothly depend on internal
coordinates. In the $(R_1,R_2)$ plane, the 
off-diagonal diabatic coupling stays localized in the inner 
region. In contrast, the coupling along bending coordinate is
delocalized across a substantial $\alpha_{\rm OCO}$ range 
in the $(R_1,\alpha_{\rm OCO})$ plane. As a result, the 
angular shape of the diabatic potentials is 
distorted compared to the adiabatic case: Diabatic potentials 
along the coupling mode $\alpha_{\rm OCO}$ are close  to the average
adiabatic potential $\frac{1}{2}(V_i^a+V_j^a)$.

\section{Conclusions}
\label{concl}

This paper describes properties of global PESs of the 
first six singlet electronic
states of CO$_2$ constructed from about 5000 symmetry unique 
ab initio points calculated with the d-aug-cc-pVQZ basis set 
using the MRD-CI method. The main results can be summarized as follows: 

\begin{enumerate}

\item Calculations accurately reproduce the known benchmarks for all
  states  and establish missing benchmarks for the future
  calculations: Bond distances and bond angles are accurate to
  within  0.1\%, known fundamental frequencies (mainly ground state
  $\tilde{X}^1\!A'$) are accurate to within 1.5\,cm$^{-1}$, the
  accuracy of the vertical
  excitation energies is expected to be better than 0.05\,eV;
  dissociation  energies agree with experimental thresholds 
  within 0.15\,eV for four covered arrangement channels. 

\item Local equilibria are abundant in the calculated states. 
  Bent OCO isomer is found in the adiabatic states 
  $1,3^1\!A'$ and $1,3^1\!A''$.
  Linear COO is found in $1,2^1\!A'$ and $1,2^1\!A''$.
  Their diabatic electronic origin    
  is clarified, and the properties, including equilibrium
  geometries, excitation energies, and vibrational
  frequencies, are established.

\item Near degeneracies can be found for each pair of six
  valence states, at linear or bent
  geometries, or at both. Avoided crossings and conical and glancing 
intersections literally 
  shape the observed topography of the excited electronic
  states. Detected intersections are not limited to the valence
  manifold and the search for electronic origins of local
  minima and barriers involves 
  valence/Rydberg and Rydberg/Rydberg intersections. 

\item Characteristic for state intersections in CO$_2$, both
  conical and glancing, is that they include several states. In the
  FC region, a fivefold intersection between $1^1\Sigma_u^-$,
  $1^1\Pi_g$, and $1^1\Delta_u$ states is found. The seam of this
  intersection forms a closed loop, spectroscopic manifestations of
  which are discussed in paper II. Outside the FC region
  at linearity, six- and sevenfold intersections are predicted, some
  of which persist over extended angular range in the bent molecule.

\item Diabatic $6\times 6$ potential matrix, with all elements smoothly
  depending on internal coordinates, is constructed using two-step local
  diabatizations of linear and bent conical intersections. 

\end{enumerate}

It is tempting to try to infer the course of photodissociation and
the principal features of the absorption spectrum --- 
the outcome of a complicated quantum mechanical calculation --- from
the constructed PESs alone. Two issues have to be resolved if one
deals with five interacting states. The first is the strength of
diabatic (intra-symmetry) and RT (inter-symmetry) coupling. If the
off-diagonal vibronic coupling is weak, the diabatic potentials should
be chosen as \lq zeroth order guides'. If the vibronic coupling is
strong, it is the adiabatic description which becomes relevant --- and
the difference between the two pictures is striking, especially along
the bending angle as Figs.\ \ref{r1gaA} and \ref{2ddia2} demonstrate. 
As discussed in paper II, the vibronic coupling is strong, the
RT interaction between $A'$ and $A''$ states is to a large
extent quenched, and the adiabatic 
potentials can be used for qualitative analysis. The global minima of 
states $2^1\!A'$ or $1^1\!A''$ are bent and lie 
$\sim 2.0$\,eV below the dissociation threshold (\ref{diss-chan1}) 
or $\sim 4.5$\,eV below the FC point. The route connecting the FC
region with these bent equilibria is barrierless. In contrast, there
is a barrier to dissociation near linearity --- 
the leftover of the lower cone of the $\Pi/\Delta$ CI. Thus, 
one expects the low energy bands in the 
absorption spectrum, associated with $2^1\!A'$ and $1^1\!A''$ states,
to reflect highly excited bending motion. This interpretation is
commonly given in
the literature:\cite{RMSM71} Instead of dissociating directly, the molecule
bends first. With growing photon energy, the contribution of
direct dissociation through linear geometries will certainly grow,
because the barrier is only about
0.2\,eV high and is located $\sim 0.4$\,eV below the FC point. Above
$\sim 9$\,eV, the valence states $3^1\!A'$ and $3^1\!A''$ will
contribute to the observed spectrum. These \lq linear' states are 
separated from the dissociation asymptote by a $\sim 1$\,eV high and
broad barrier, with the implication that CO$_2$ in these states can
decay only through non-adiabatic interactions with the
dissociation continuum of the lower states. In other words, one
expects to see a resonance-dominated absorption spectrum. The next
qualitative change in the absorption 
spectrum within the valence manifold can be expected
after the photon energy reaches the top of the dissociation barrier in
the upper valence states and allows direct dissociation from linearity.

The above discussion is based on 1D and 2D potential cuts and the data in Table
\ref{table3} --- given the adiabatic representation is the
adequate one. However, there is another important piece of information
still missing, namely the TDMs with the ground state. As has already
been mentioned in Sect.\ \ref{ex1}, the electronic transitions in the wavelength
range of 160\,nm --- 120\,nm are forbidden. The bands are observed
only because the TDMs are not constant but strongly change with
molecular coordinates as one moves away from the high-symmetry
FC point. This dependence is a manifestation of the
Herzberg-Teller effect\cite{HERZBERG67} which 
plays the leading role in shaping the
absorption bands, is at least as important as the potential
profiles discussed above, and has to be considered on equal footing
with the state intersections. The discussion of the coordinate
dependence of the TDMs is deferred to
paper II.



\acknowledgments{ 
Financial support by the  Deutsche
  Forschungsgemeinschaft is gratefully acknowledged
}
\clearpage


\begin{thebibliography}{10}
\bibitem{NOTE-CO23-1}
{\rm CASSCF energies at $R_1 = 2.4\,a_0$ , $R_2 = 5.0\,a_0$, and $\alpha_{\rm
  OCO} = 179^\circ$, relative to the minimum of the $\tilde{X}^1\!A'$ curve,
  are $E_e \approx 7.1$\,eV, 8.9\,eV, and 11.2\,eV for channels (3), (4), and
  (7), respectively.}
\bibitem{NOTE-CO23-2}
{\rm Note that the avoided Rydberg/valence crossings between $1,2^1\Delta_u$
  and $1,2^1\Sigma_u^-$ states, investigated in Ref. \onlinecite{SFCRW91}, lie
  outside the cuts shown in Fig.\ 11.}
\bibitem{NOTE-CO21-3}
{\rm  Whenever \lq diabatic representation' is mentioned, a quasi-diabatic
  representation,\cite{K04} in which the kinematic singularities and the
  potential cusps at the intersections are removed, is meant.}
\end{thebibliography}

\clearpage

\begin{table}
\caption{
Properties of the adiabatic ground electronic state 
$\tilde{X}^1\Sigma_g^+$: 
The OCO bond angle ($\alpha_{\rm OCO}$, in $^\circ$),
CO and OO bond distances (in $a_0$),  
and the energy $\Delta E$ relative to the global minimum (in eV, 
including the vibrational ZPE corrections),
calculated for three equilibrium geometries and the two arrangement
channels. 
}
\label{table1}
\begin{ruledtabular}
\begin{tabular}{lcccccc}
Geometry & $\alpha_{\rm OCO}$ & $R_{\rm CO,1}$ & $R_{\rm CO,2}$ &  
$R_{\rm OO}$ & $\Delta E$\tablenotemark[1]  & Reference \\
\hline
OCO linear & 180.0 & 2.1991 & 2.1991 & 4.3982 & 0.0 & this work \\
           & 180.0 & 2.2050 & 2.2050 & 4.4100 & 0.0 &  \onlinecite{HM00A}\\
           & 180.0 & 2.1924 & 2.1924 & 2.3848 & 0.0 
&  \onlinecite{KRW88,SFCCRWB92}\\
           & 180.0 & 2.1960 & 2.1960 & 4.3920 & 0.0 & 
\onlinecite{NISTDATABASE1}\tablenotemark[2] \\
           &       &        &        &        &     &                   \\
OCO bent   & 73.2 & 2.51 & 2.51 & 2.99 & 5.90 & this work \\
           & 72.9 & 2.52 & 2.52 & 2.99 & 6.04& \onlinecite{HM00A} \\
           & 73.1 & 2.53 & 2.53 & 3.01 & 5.97\tablenotemark[3]  
& \onlinecite{XR94} \\
           &      &        &        &        &     &                   \\
COO linear & 0.0  & 2.20 & 4.65 & 2.45 & 7.35\tablenotemark[3]   & this work \\
           & 0.0  & 2.20 & 4.65 & 2.45 & 7.14  &  \onlinecite{HM00A} \\
           & 0.0  & 2.17 & 5.37 & 3.20 & 6.86 &  \onlinecite{XR94} \\
           &      &        &        &        &     &                   \\
O$(^1D)$+CO$(\tilde{X}^1\Sigma^+)$ & ---  & 2.14 & $\infty$   & $\infty$ 
& 7.28 & this work \\
                                   &   & 2.175 & $\infty$   & $\infty$ 
& 7.64 & \onlinecite{HM00A} \\
                                   &   & 2.132 & $\infty$   & $\infty$ 
& 7.41 & \onlinecite{HERZBERG67,AMS79}\tablenotemark[2] \\
           &      &        &        &        &     &                   \\
C$(^3P)$+O$_2(\tilde{X}^3\Sigma_g^-)$ & ---  & $\infty$ & $\infty$   
& 2.30 & 11.35 & this work \\
                                      &      & $\infty$ & $\infty$   
& 2.356 & 11.49 & \onlinecite{HM00A} \\
                                      &      & $\infty$ & $\infty$   
& 2.283 & 11.52 & \onlinecite{HERZBERG67}\tablenotemark[2] \\
           
\end{tabular}
\end{ruledtabular}
\tablenotetext[1]{The ab initio ZPEs are: 
ZPE(CO$_2$)\,=\,0.314\,eV, ZPE(CO)\,=\,0.130\,eV, and 
ZPE(O$_2$)\,=\,0.095\,eV.}
\tablenotetext[2]{Experimental values.}
\tablenotetext[3]{Energy without ZPE correction.}
\end{table}
\newpage
\clearpage

\begin{table}
\caption{Energies (in cm$^{-1}$) of the low lying vibrational states 
in the ground electronic state $\tilde{X}^1\Sigma_g^+$, 
measured relative to the ground level $(0,0^0,0)$. The states calculated
with the original ab initio potential (\lq PES1') and with the potential 
rescaled along the symmetric stretch and the bend (\lq PES2') are compared
with the best theoretical estimates of Refs.\ 
\onlinecite{RHYHTST07,YHH08} and with the
experimental values of Ref.\ \onlinecite{C79A} (denoted \lq Exp'). 
The energy difference $E_{\rm observed} - E_{\rm calculated}$
(in cm$^{-1}$) is shown in parenthesis. The quantum numbers
$(v_s,v_b^l,v_a)$ are defined in text. 
 }
\label{table2}
\begin{ruledtabular}
\begin{tabular}{lcccc}
State & Refs.\ \onlinecite{RHYHTST07,YHH08} & PES1 & PES2 & Exp \\
\hline
$(0,0^0,0)$\tablenotemark[1] & 0.0 & 0.0 & 0.0& 0.0 \\
$(0,1^1,0)$ & 669.1 (-1.7) & 668.6 (-1.2) & 670.3 (-2.9)& 667.4 \\
$(1,0^0,0)$ & 1288.9 (-3.5) & 1265.3 (20.1) & 1284.1 (1.3)& 1285.4 \\
$(0,2^2,0)$ & 1339.6 (-4.5) & 1336.6 (-1.5) & 1336.8 (-1.7)& 1335.1 \\
$(0,2^0,0)$ & 1389.3 (-1.1) & 1373.5 (14.7) & 1389.2 (-1.0)& 1388.2 \\
$(1,1^1,0)$ & 1938.0 (-5.5) & 1913.2 (19.3) & 1933.4 (-0.9)& 1932.5 \\
$(0,3^3,0)$ & 2011.4 (-8.2) & 2005.9 (-2.7) & 2006.1 (-2.9)& 2003.2 \\
$(0,3^1,0)$ & 2080.0 (-3.1) & 2061.7 (15.2) & 2079.6 (-2.7)& 2076.9 \\
$(0,0^0,1)$ & 2349.2 (0.0) & 2350.6 (-1.4) & 2351.4 (-2.2)& 2349.2 \\
$(1,2^0,0)$ & 2552.0 (-8.6) & 2516.4 (27.0) & 2549.5 (-6.1)& 2543.4 \\
$(2,0^0,0)$ & 2676.3 (-5.2) & 2626.0 (45.1) & 2668.3 (2.8)& 2671.1 \\
$(0,4^0,0)$ & 2809.1 (-12.0) & 2761.8 (35.3) & 2790.2 (6.9)& 2797.1 \\
$(1,2^2,0)$ & 2589.8 (-4.7) &       & 2581.6 (3.5) & 2585.1\\
\end{tabular}
\end{ruledtabular}
\tablenotetext[1]{The energy difference between the ground vibrational
level and the potential minimum is 
2516\,cm$^{-1}$ for PES1 and 2533\,cm$^{-1}$ for PES2.
The ZPE\,=\,2508.5\,cm$^{-1}$, given in the 
NIST database,\cite{NISTDATABASE1} 
is evaluated from the fundamental frequencies via
$\omega_b +  \frac{1}{2}(\omega_s^0 + \omega_a)$. 
For PES1 and PES2, this value is 2503.6\,cm$^{-1}$
and 2512.5\,cm$^{-1}$, respectively.
}
\end{table}
\newpage
\clearpage

\begin{table}
\caption{Properties of the PESs of the first five excited singlet states 
of CO$_2$: Vertical excitation energy $T_v$ (in eV); band origin
$T_0$ (in eV), which includes ZPEs of the ground
and the excited electronic states; 
equilibrium CO bond lengths $R_{1,e}$ and $R_{2,e}$ (in $a_0$);
equilibrium OCO bond angle $\alpha_e$ (in $^\circ$); quantum mechanical 
vibrational frequencies $\omega_{s}$, $\omega_{a}$, and $\omega_{b}$ 
near equilibrium
(in cm$^{-1}$);  quantum
mechanical dissociation thresholds $D_0({\rm O/CO})$ and $D_0({\rm C/O2})$
in the O+CO and C+O$_2$ arrangement channels, respectively (in eV). }
\label{table3}
\begin{ruledtabular}
\begin{tabular}{lccccc}
$C_s$ 
& $1^1\!A''$\tablenotemark[1] & $2^1\!A'$\tablenotemark[1] & $2^1\!A''$ &
$3^1\!A'$\tablenotemark[2]& $3^1\!A''$\tablenotemark[2] \\
$D_{\infty h}$ & $^1\Pi_g$ & $^1\Pi_g$ &
$^1\!\Sigma_u^-$ & $^1\Delta_u$ & $^1\!\Delta_u$ \\
O+CO channel\tablenotemark[3] & $^1\!D/^1\Sigma$ & $^1\!D/^1\Sigma$ &
$^3\!P/^3\Pi$ & $^1\!D/^1\Sigma$ & $^1\!D/^1\Sigma$ \\
C+O$_2$ channel\tablenotemark[3]  & $^1\!D/^1\Delta$ & $^1\!D/^1\Delta$ &
$^1\!D/^1\Delta$ & $^3\!P/^3\Sigma$& $^3\!P/^3\Sigma$ \\
\hline
$T_0$ &5.36 & 5.39 & 7.95 & 8.67 & 8.67\\
$T_v$  & 8.92 & 8.92 & 8.79 & 9.16 & 9.17 \\ 
$T_v$, Refs.\,\onlinecite{SFCCRWB92} & 9.00 & 9.00 & 9.19 & 9.28 & 9.28 \\ 
$R_{1,e} $ & 2.37 & 2.36
& 2.41 & 2.25 & 2.25 \\
$R_{2,e} $ & 2.37 & 2.36 & 2.41 & 2.80 & 2.80\\
$R_{e}$, Ref.\,\onlinecite{KRW88,SFCCRWB92} 
& 2.38 & 2.38 & 2.40 & 2.29 & 2.29\\
$\alpha_{e} $ & 127.3 & 117.9 & 176.0 & 180.0 & 180.0 \\ 
$\alpha_{e}$, Ref.\,\onlinecite{SFCCRWB92} 
& 127.0 & 117.8 & 180.0 & 180.0 & 180.0 \\ 
$\omega_{s}$ & 1283 & 1340 & 1015& 520& 550\\
$\omega_{a}$ & 905 & 865 & 1118& 1560 & 1550\\
$\omega_{b}$ & 670 & 580 & 577& 2290 & 3100\\
$D_0({\rm O/CO})$\tablenotemark[4]  & 7.27& 7.27& 11.31& 7.27 &7.27\\
$D_0({\rm C/O2})$\tablenotemark[4]  & 11.34& 11.34& 13.59& 13.59 &13.59\\
\end{tabular}
\end{ruledtabular}
\tablenotetext[1] {Experimental data from Ref.\
  \onlinecite{SFCCRWB92} are: For $1^1\!A''$ $R_e = 2.28 \pm
  0.02\,a_0$, $\alpha_e = 129 \pm 1^\circ$, $T_0 \le 6.2$\,eV,
  $\nu_b \approx 632$\,cm$^{-1}$; for $2^1\!A'$ $R_e = 2.35 \pm
  0.015\,a_0$, $\alpha_e = 122 \pm 2^\circ$, $T_0 \approx 5.7$\,eV,
$\nu_b \approx 600$\,cm$^{-1}$.}
\tablenotetext[2]{For the $3^1\!A'$ and $3^1\!A''$ states, the
  geometries refer to the local minimum in the FC region. The
vibrational frequencies are strongly perturbed by CI cusps, and
the corresponding ZPEs are omitted in $T_0$.}
\tablenotetext[3]{Dissociation thresholds, labeled with 
  electronic states of atomic/diatomic fragment,
are correlated with the diabatic states of CO$_2$.}
\tablenotetext[4]{Experimental dissociation thresholds are given in Eqs.\
  (\ref{diss-chan1}), (\ref{diss-chan3}), (\ref{diss-chan2}), and
(\ref{diss-chan4}). Ab initio ZPEs are: ZPE[CO$_2(\tilde{X}^1\Sigma_g^+$)] = 
0.314\,eV; ZPE[CO$(\tilde{X}^1\Sigma^+)$] = 0.130\,eV;  ZPE[CO$(a^3\Pi)$] =
0.099\,eV; ZPE[O$_2(\tilde{X}^3\Sigma_g^-)$] = 0.095\,eV; 
ZPE[O$_2(^1\Delta_g)$] = 0.080\,eV.}
\end{table}

\newpage
\clearpage

\begin{figure}[t]
\caption{Cuts through the PESs
of the states $1,2,3^1A'$ (a,c,e) and $1,2,3^1A''$ 
(b,d,f) along one CO bond distance $R_{\rm CO}$. 
The second CO bond is fixed at $2.2\,a_0$; the fixed bond angle is
indicated in each panel.   
Dots are the ab initio adiabatic energies. Solid lines are
states diabatized at bent CIs as described in Sect.\ \ref{qd1}. 
For $A''$ states,
the ground electronic state $\tilde{X}^1\!A'$, shown with gray dots, sets the
vertical energy scale. At linearity, components of the same orbitally
degenerate state $\Pi_g$ or $\Delta_u$
have the same color in panels (a) and (b).  
\label{curco1}
}
\end{figure}

\begin{figure}[t]
\caption{Cuts through the PESs 
of the states $1,2,3^1A'$ (a,c,e) and $1,2,3^1A''$ 
(b,d,f) along one CO bond distance $R_{\rm CO}$. The fixed bond angle is
indicated in each panel. The second CO bond 
is fixed at $2.2\,a_0$ in (a-d). In (e) and (f), the OO
bond is kept fixed at 2.3$\,a_0$. 
Dots are the ab initio adiabatic energies. Solid lines are
states diabatized at bent CIs as described in Sect.\ \ref{qd1}. 
For $A''$ states,
the ground electronic state $\tilde{X}^1A'$, shown with gray dots, sets the
vertical energy scale. 
\label{curco2}
}
\end{figure}

\begin{figure}[t]
\caption{Cuts through the PESs
of the states $1,2,3^1A'$ (a,c,e) and $1,2,3^1A''$ 
(b,d,f) along bond angle $\alpha_{\rm OCO}$. The two 
CO bonds are fixed, one 
at $2.2\,a_0$ and the other as indicated in each panel. 
Dots are the ab initio adiabatic energies. Solid lines are
states diabatized at bent CIs as described in Sect.\ \ref{qd1}. 
\label{curwink1}
}
\end{figure}

\begin{figure}[t]
\caption{Cuts through the PESs
of the states $1,2,3^1A'$ (a,c,e) and $1,2,3^1A''$ 
(b,d,f) along bond angle $\alpha_{\rm OCO}$. The difference with
Fig.\ \ref{curwink1} is that one CO bond is fixed at 2.5\,$a_0$. 
\label{curwink2}
}
\end{figure}

\begin{figure}[t]
\caption{Contour maps 
of the ground electronic state $\tilde{X}^1\!A'$: $(R_1,R_2)$ plane, with
$\alpha_{\rm OCO}$ fixed at 179$^\circ$ (a) and $70^\circ$ (b), and 
$(R_1,\alpha_{\rm OCO})$ plane with $R_{\rm CO}$ fixed at 
$2.2\,a_0$ (c) and $2.6\,a_0$ (d). Energy of the 
dotted contour in (a), (c), and (d) is 2.1\,eV, in (b) it is 8.4\,eV. Contour
spacing is 0.35\,eV. 
\label{2dx}
}
\end{figure}

\begin{figure}[t]
\caption{Contour maps in the $(R_1,R_2)$ plane 
of the adiabatic states $2^1\!A'$ (a) and 
$3^1\!A'$ (b) for $\alpha_{\rm OCO}=179^\circ$. The states are 
components of  $1^1\Pi_g$ and $1^1\Delta_u$, respectively. 
 Black solid line in each panel indicates the closed CI seam for 
$A'$ (a) and $A''$ (b) symmetry states. Black dashed line in (a) shows the 
path along which the matrix elements $\left|\langle
  iA''|\hat{L}_z|jA'\rangle\right|$ in Fig.\ \ref{mrcilz} are calculated.
Energy of the dotted contour is 7.5\,eV, and the contour
spacing is 0.25\,eV. 
\label{r1r2A}
}
\end{figure}

\begin{figure}[t]
\caption{MRD-CI matrix elements of $\hat{L}_z$, 
$\left|\langle iA''|\hat{L}_z|jA'\rangle\right|$,  for electronic state pairs
$(i=1,j=2)$ (black), $(i=2,j=2)$ (purple), $(i=3,j=2)$ (brown), 
and $(i=3,j=3)$ (red) calculated as functions of the CO bond distance
$R_1$ for $R_2 = 2.2\,a_0$ and $\alpha_{\rm OCO}=179^\circ$ 
along the dashed line shown in Fig.\ \ref{r1r2A} 
\label{mrcilz}
}
\end{figure}

\begin{figure}[t]
\caption{Contour maps of the adiabatic state $2^1A''$:
$(R_1,R_2)$ plane, with
$\alpha_{\rm OCO}$ fixed at 179$^\circ$ (a) and $70^\circ$ (b), and 
$(R_1,\alpha_{\rm OCO})$ plane with $R_{\rm CO}$ fixed at 
$2.2\,a_0$ (c) and $2.6\,a_0$ (d).
Energy of the 
dotted contour in (a), (c), and (d) is 9.0\,eV, in (b) it is 12.5\,eV. Contour
spacing is 0.25\,eV. 
\label{2d2app}
}
\end{figure}

\begin{figure}[t]
\caption{Contour maps in the $(R_1,\alpha_{\rm OCO})$ plane 
of the adiabatic states $2^1\!A'$ (a), $3^1\!A'$ (b), $1^1\!A''$ (c),
and $3^1\!A''$ (d).  The second CO
bond is fixed at 2.2$\,a_0$. Energy of the dotted contour is 7.6\,eV, and the 
contour spacing is 0.20\,eV. 
\label{r1gaA}
}
\end{figure}

\begin{figure}[t]
\caption{Contour maps in the $(R_1,R_2)$ plane 
of the adiabatic states $2^1\!A'$ (a), $3^1\!A'$ (b), $1^1\!A''$ (c),
and $3^1\!A''$ (d).  The angle $\alpha_{\rm OCO}$ is fixed at $160^\circ$. 
Energy of the dotted contour is 8.0\,eV, and the 
contour spacing is 0.25\,eV. 
\label{r1r2B}
}
\end{figure}

\begin{figure}[t]
\caption{CASSCF energies of  the 
electronic states 1---5$^1\!A'$ (a,c,e) and 1---5$^1\!A''$ 
(b,d,f) along one CO bond distance $R_{\rm CO}$. The second CO bond is
fixed at $2.4\,a_0$: the fixed bond angle is indicated in each
panel. In all panels, dots and solid lines 
denote adiabatic energies. In panels (a) and (b), lines are color
coded according to the ab initio $\langle L_z^2 \rangle$ values
as explained in text. Following a particular color,
one follows a diabatic state. Diabatic assignments using spectroscopic
symbols are given. Arrows in panel (b) indicate the
$\Delta/\Pi$ CIs discussed in Sect.\ \ref{ex2}. For $A''$ states,
the ground electronic state $\tilde{X}^1\!A'$, shown with gray dotted line, 
sets the vertical energy scale. 
\label{mcscfr1}
}
\end{figure}

\begin{figure}[t]
\caption{Adiabatic CASSCF energies of  
states 1---5$^1\!A'$ (a) and 1---5$^1\!A''$ (b) along the bond angle.
Fixed CO bond distances are $2.4\,a_0$  and $2.5\,a_0$. The two
uppermost states in each symmetry block are drawn gray. 
Arrow indicates direction towards the carbene-like OCO
minimum in the given state. 
Correlation with the diabatic states at linearity is marked to
the right of each panel. 
\label{mcscfw1}
}
\end{figure}

\begin{figure}[t]
\caption{Contour maps in the $(R_1,\alpha_{\rm OCO})$ plane 
of the adiabatic states $2^1\!A'$ (a), $3^1\!A'$ (b), $1^1\!A''$ (c),
and $3^1\!A''$ (d).  The second CO
bond is fixed at 2.6$\,a_0$. Energy of the dotted contour is 9.0\,eV, and the 
contour spacing is 0.20\,eV. 
\label{r1gaB}
}
\end{figure}

\begin{figure}[t]
\caption{The mixing angle $\tilde{\Theta}$ 
for the states $2,3^1\!A'$
  calculated as a function of one CO bond distance.
Solid lines and open circles denote $\tilde{\Theta}$ from 
the regularized diabatic states model of Eqs.\ 
(\ref{diarep2})---(\ref{diarep3}). Solid squares are the values obtained
using quasi-diabatization procedure in MOLPRO.  The second CO bond is
fixed at $R_1 = 2.2\,a_0$; the bond angle $\alpha_{\rm OCO}$ is
indicated in each panel. The ab initio mixing angle, obtained with the
cc-pVQZ basis set, is lifted by 25$^\circ$. 
\label{theta}
}
\end{figure}

\begin{figure}[t]
\caption{Diabatic potentials (diagonal matrix elements) for the
  $\Sigma$ (green), $\Pi$
  (blue), and $\Delta$ (red) states and the off-diagonal
couplings (black solid line) as functions of one CO bond distance. 
$A'$ ($A''$) states are shown in the left (right) panels. 
The second CO bond is fixed at $2.2\,a_0$; the fixed bond angle is
indicated in each panel. 
Black dashed line indicates the ground electronic state.  
\label{1ddia}
}
\end{figure}

\begin{figure}[t]
\caption{The diabatic potentials 
$V_{\Pi'}$ (a),  $V_{\Delta'}$ (c), $V_{\Pi''}$ (b), and
$V_{\Delta''}$ (d), and the off-diagonal coupling elements 
$V_{\Pi'\Delta'}$ (e) and  $V_{\Pi''\Delta''}$ (f) in the
$(R_1,R_2)$ plane with the bond angle fixed at 175$^\circ$. 
Energy of the dotted contour in (a), (b), (c), and (d) 
is 7.5\,eV, and the contour spacing is 0.25\,eV. 
In (e) and (f), the respective numbers are -0.3\,eV and 0.05\,eV. 
\label{2ddia1}
}
\end{figure}

\begin{figure}[t]
\caption{The diabatic potentials 
$V_{\Pi'}$ (a),  $V_{\Delta'}$ (c), $V_{\Pi''}$ (b), and
$V_{\Delta''}$ (d), and the off-diagonal coupling elements 
$V_{\Pi'\Delta'}$ (e) and $V_{\Pi''\Delta''}$ (f) in the
$(R_1,\alpha_{\rm OCO})$ plane with the CO bond fixed at 2.2\,$a_0$. 
Energy of the dotted contour in (a), (b), (c), and (d) 
is 8.6\,eV, and the contour spacing is 0.20\,eV. 
In (e) and (f), the respective numbers are -1.20\,eV and 0.20\,eV. 
\label{2ddia2}
}
\end{figure}


\clearpage
\newpage
\mbox{ }
\vspace{1cm}

\includegraphics[angle=0,scale=0.7]{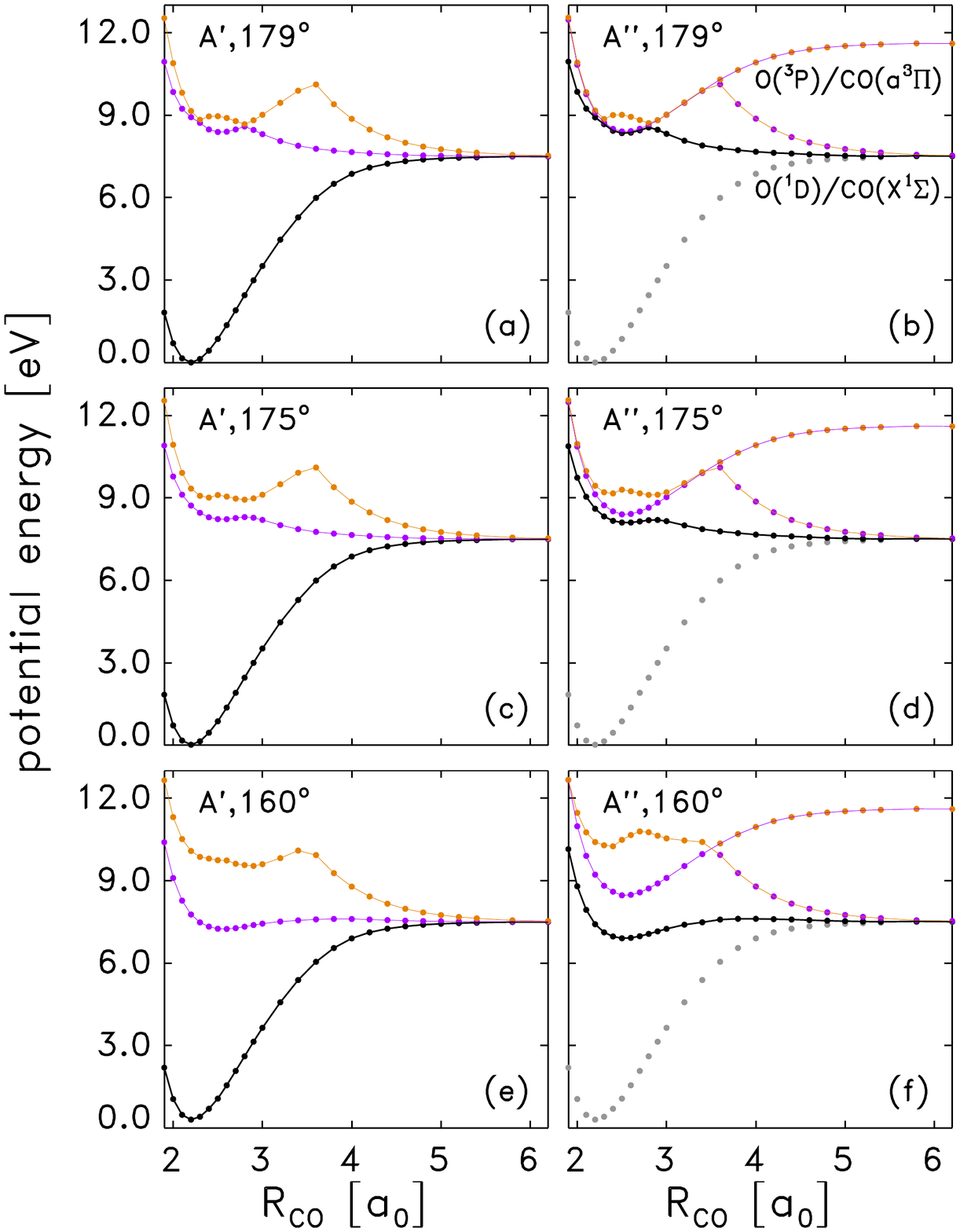}

\vspace{-1cm}

Fig.\ 1

\newpage
\mbox{ }
\vspace{1cm}

\includegraphics[angle=0,scale=0.7]{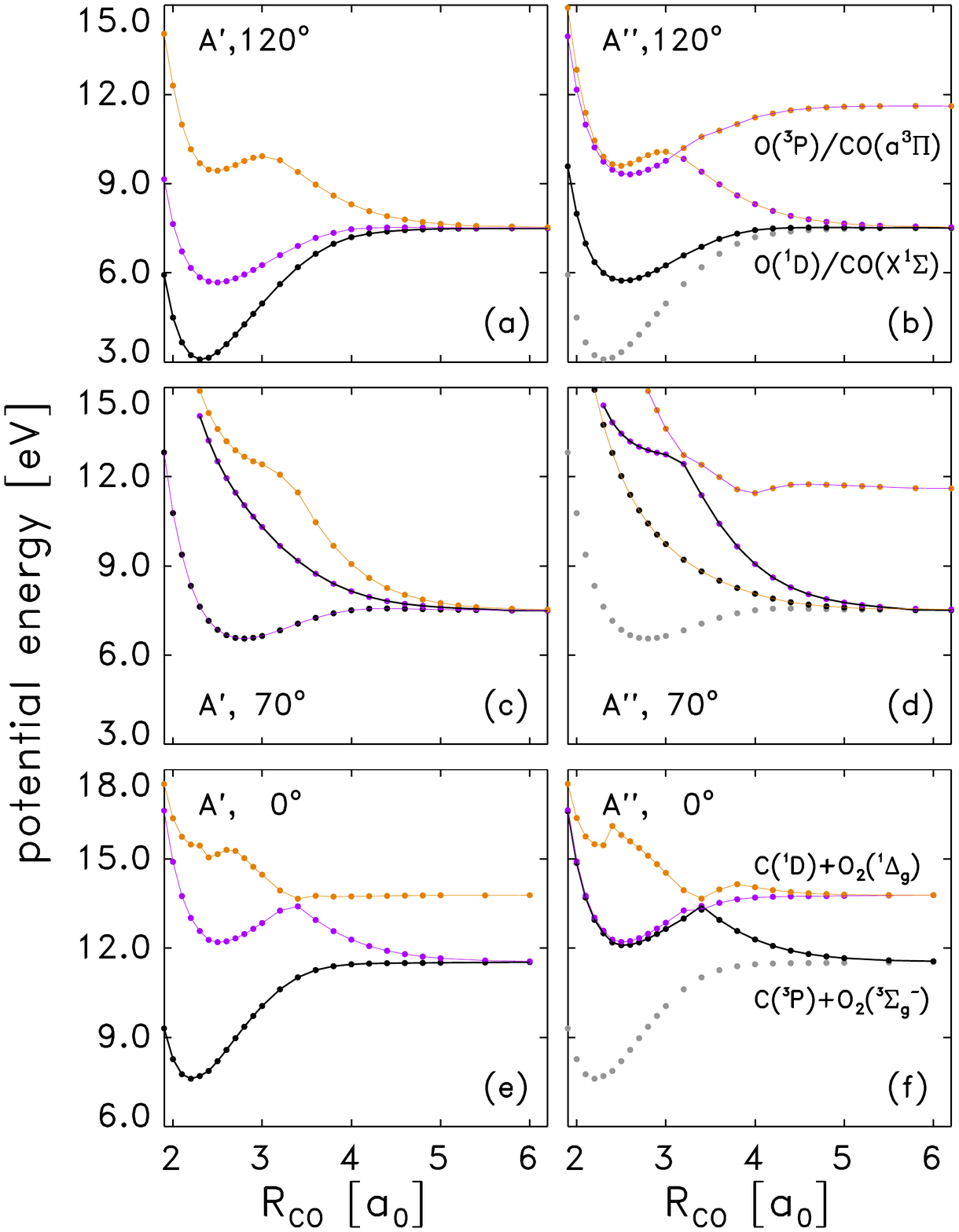}

\vspace{-1cm}

Fig.\ 2

\newpage
\mbox{ }
\vspace{1cm}

\includegraphics[angle=0,scale=0.7]{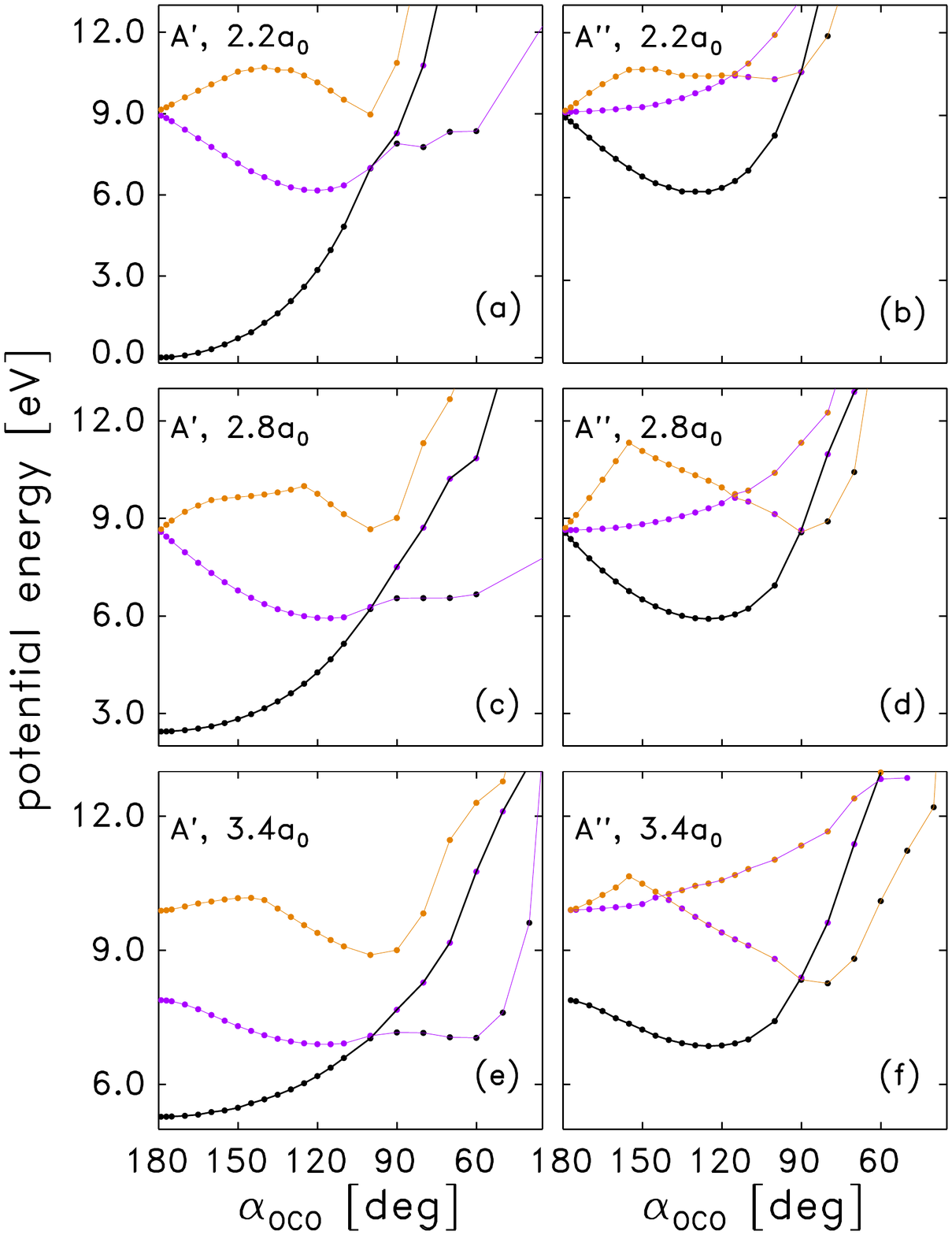}

\vspace{-1cm}

Fig.\ 3

\newpage
\mbox{ }
\vspace{1cm}

\includegraphics[angle=0,scale=0.7]{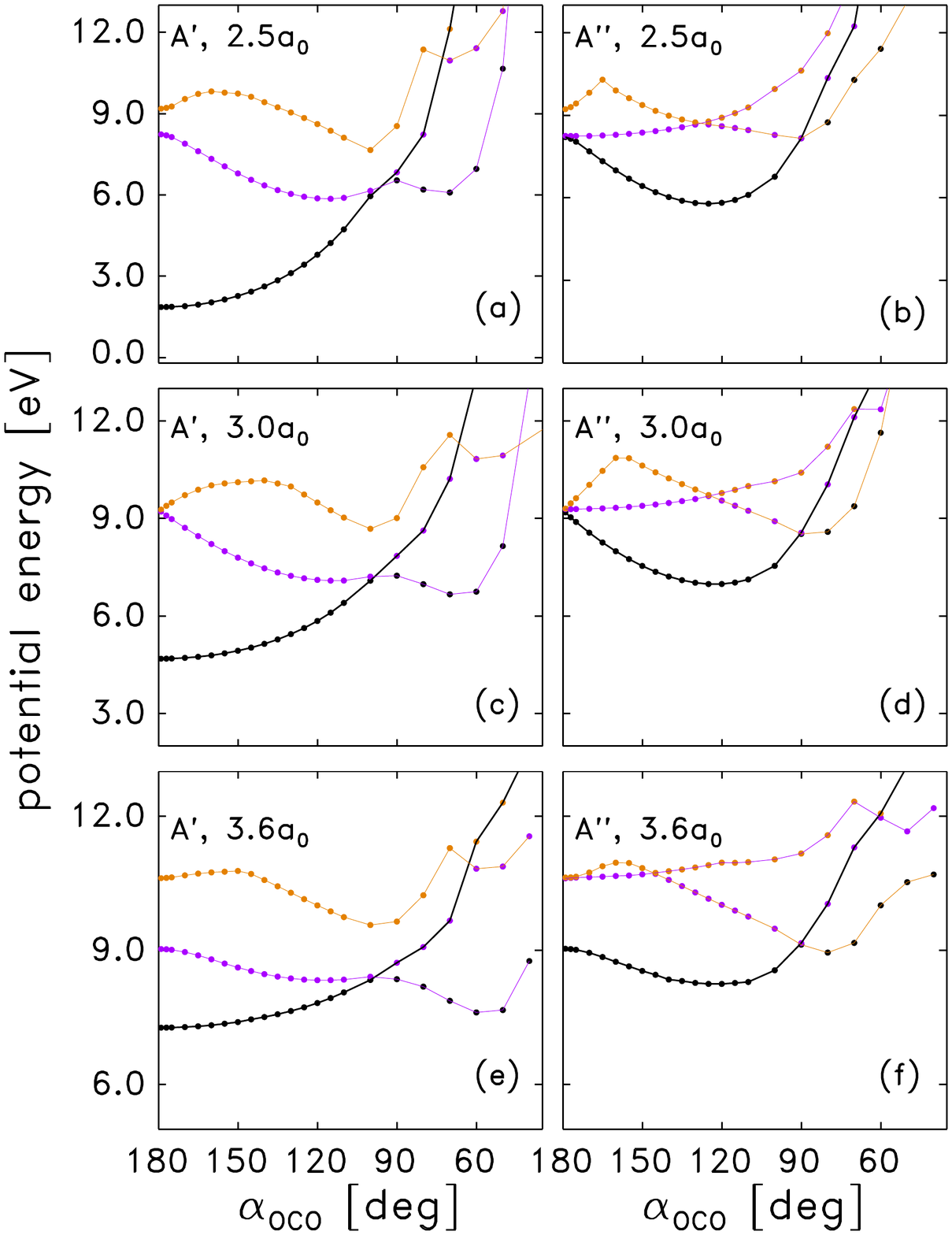}

\vspace{-1cm}

Fig.\ 4

\newpage
\mbox{ }
\vspace{-2cm}

\includegraphics[angle=0,scale=0.7]{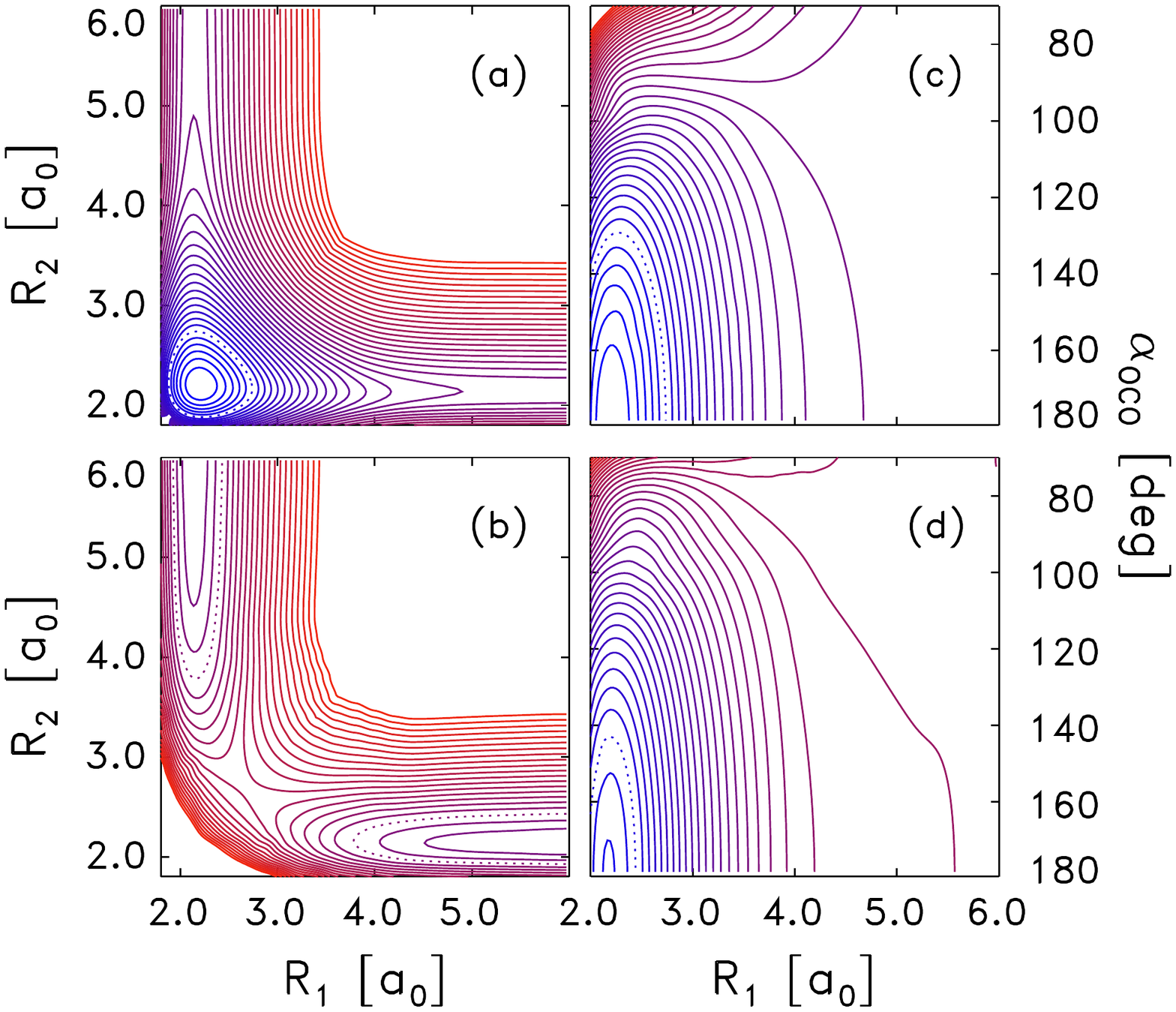}

\vspace{-1cm}

Fig.\ 5

\newpage
\mbox{ }
\vspace{-2cm}

\includegraphics[angle=0,scale=0.7]{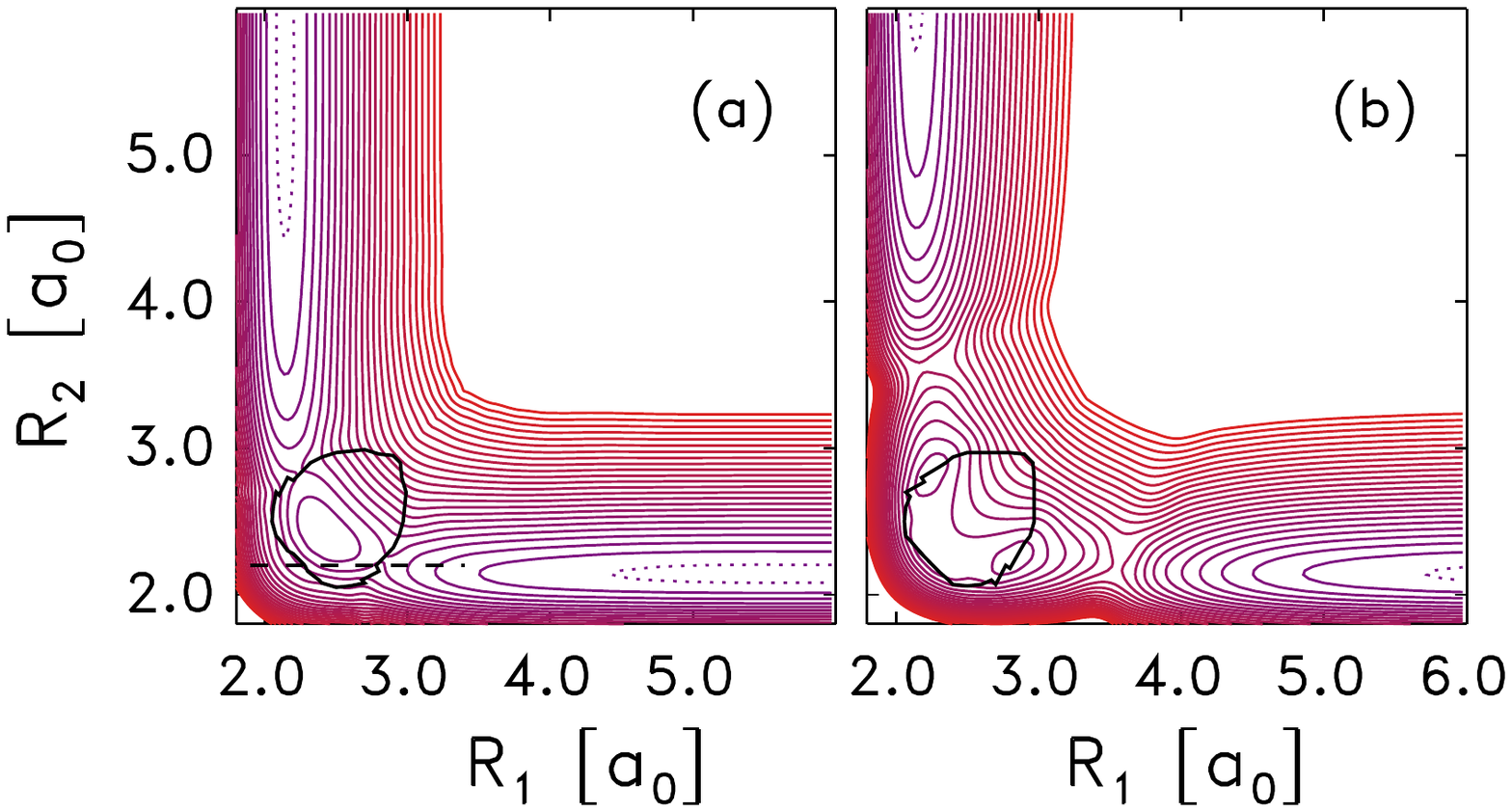}

\vspace{-1cm}

Fig.\ 6

\newpage
\mbox{ }
\vspace{1cm}

\includegraphics[angle=0,scale=0.7]{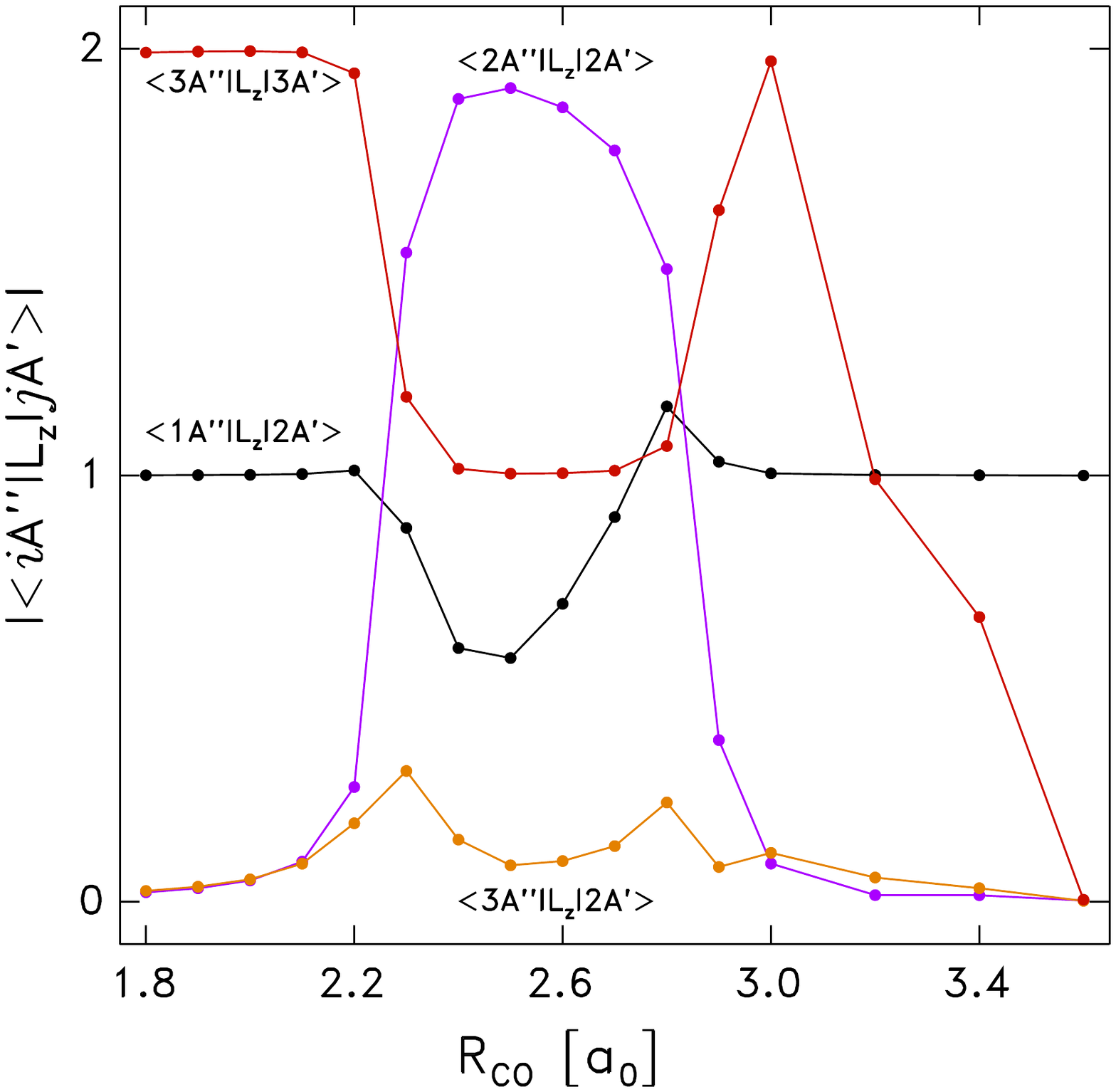}

\vspace{-1cm}

Fig.\ 7

\newpage
\mbox{ }
\vspace{-2cm}

\includegraphics[angle=0,scale=0.7]{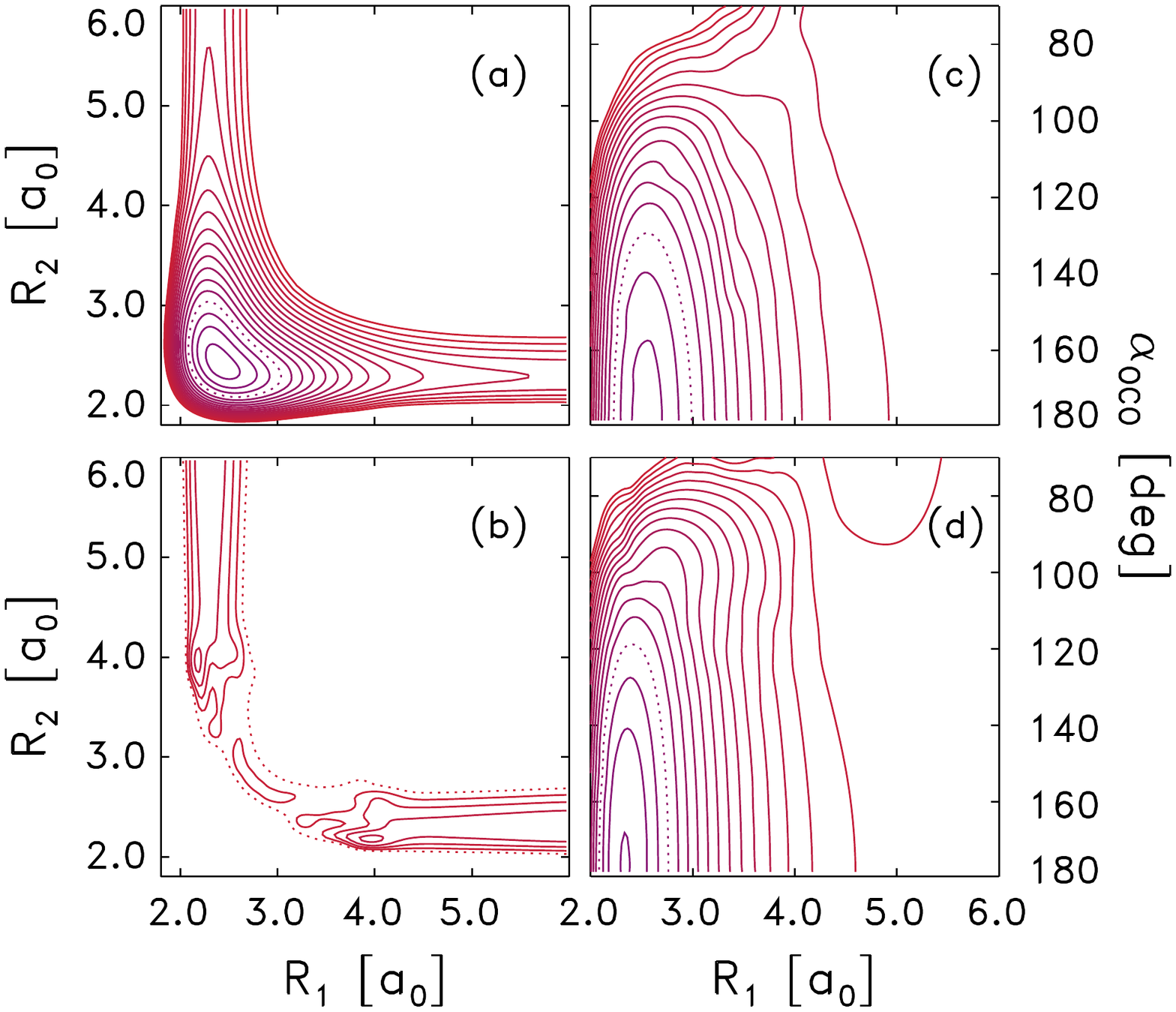}

\vspace{-1cm}

Fig.\ 8

\newpage
\mbox{ }
\vspace{-2cm}

\includegraphics[angle=0,scale=0.7]{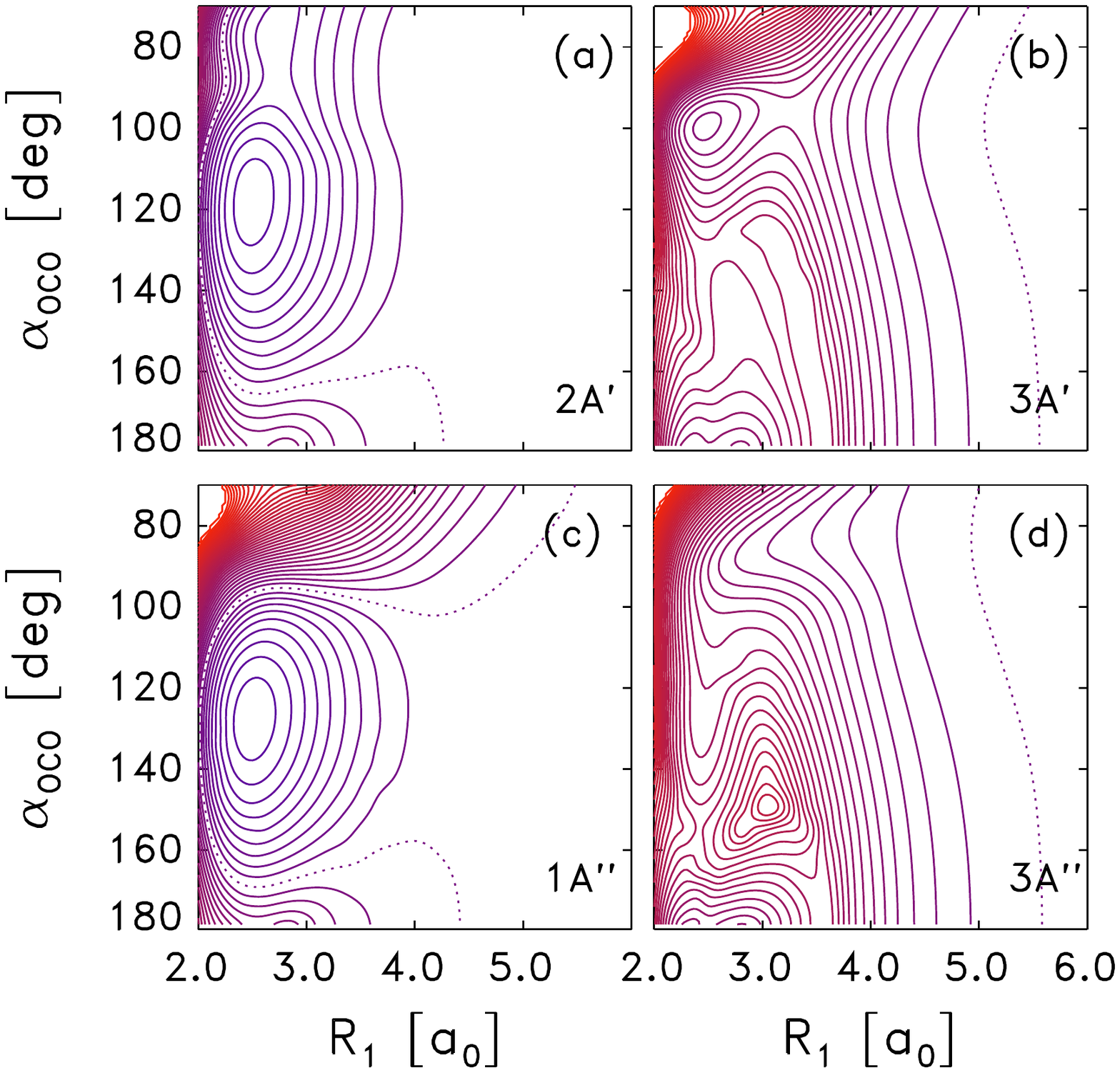}

\vspace{-1cm}

Fig.\ 9

\newpage
\mbox{ }
\vspace{-2cm}

\includegraphics[angle=0,scale=0.7]{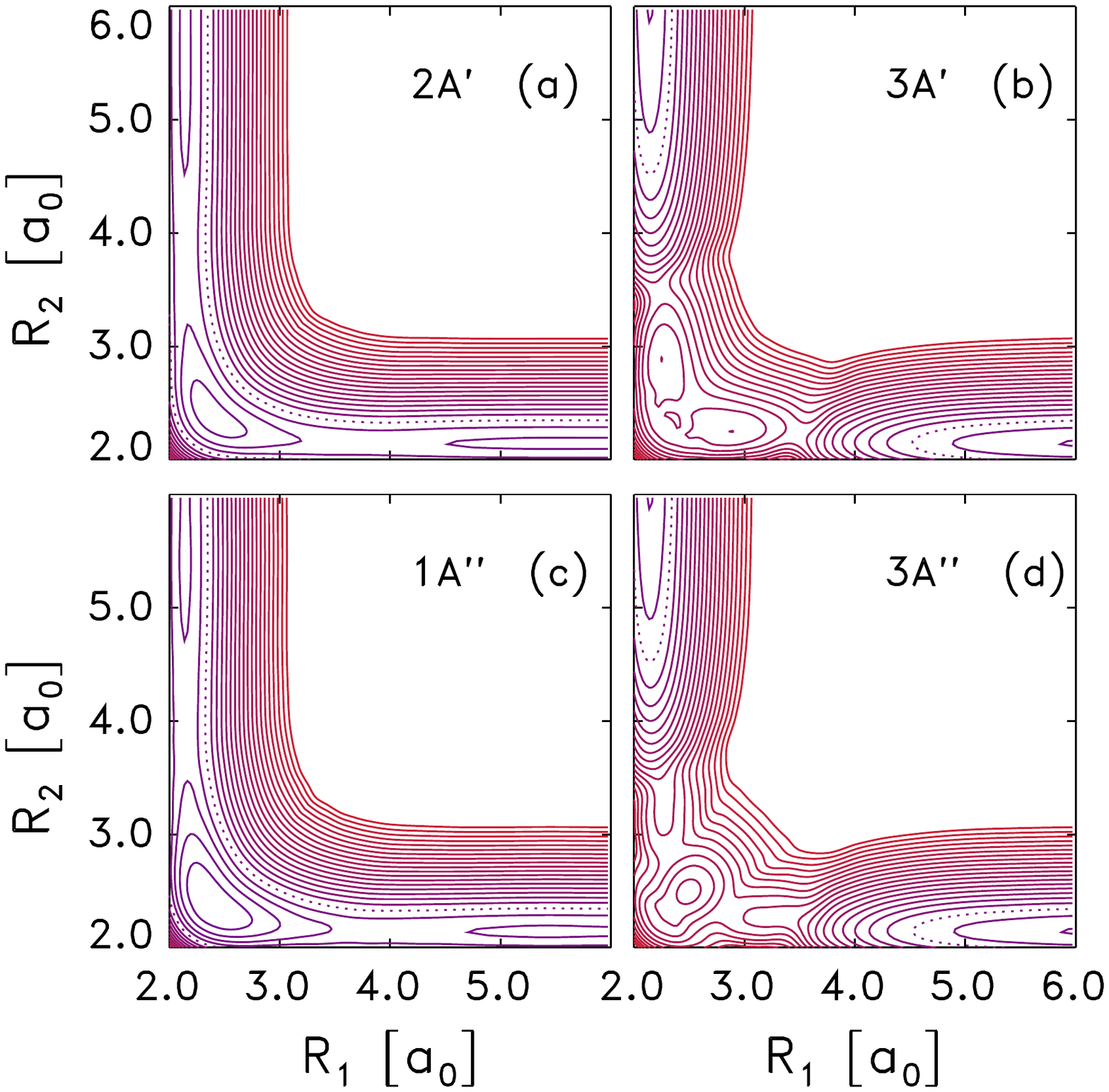}

\vspace{-1cm}

Fig.\ 10

\newpage
\mbox{ }
\vspace{1cm}

\includegraphics[angle=0,scale=0.7]{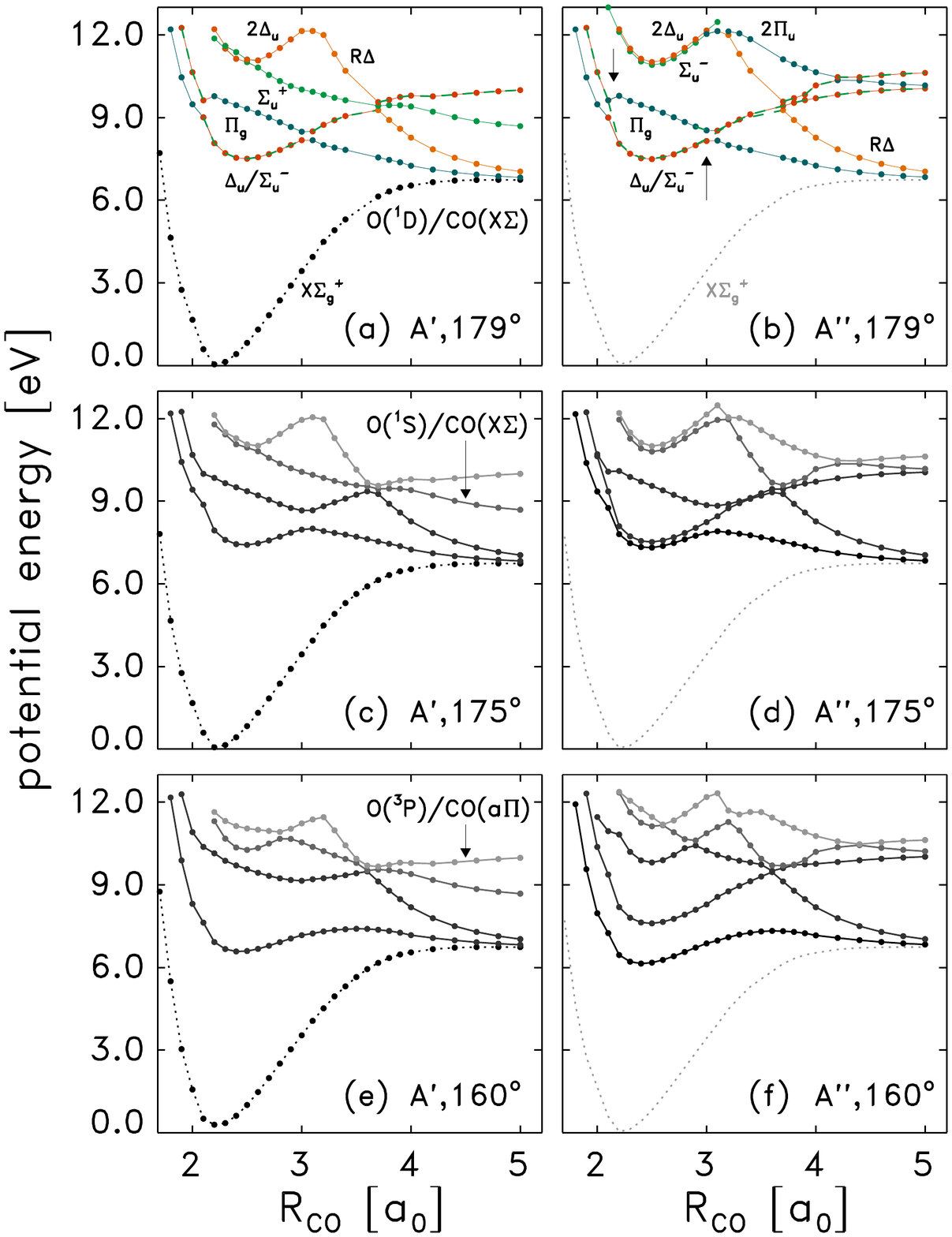}

\vspace{-1cm}

Fig.\ 11

\newpage
\mbox{ }
\vspace{1cm}

\includegraphics[angle=0,scale=0.7]{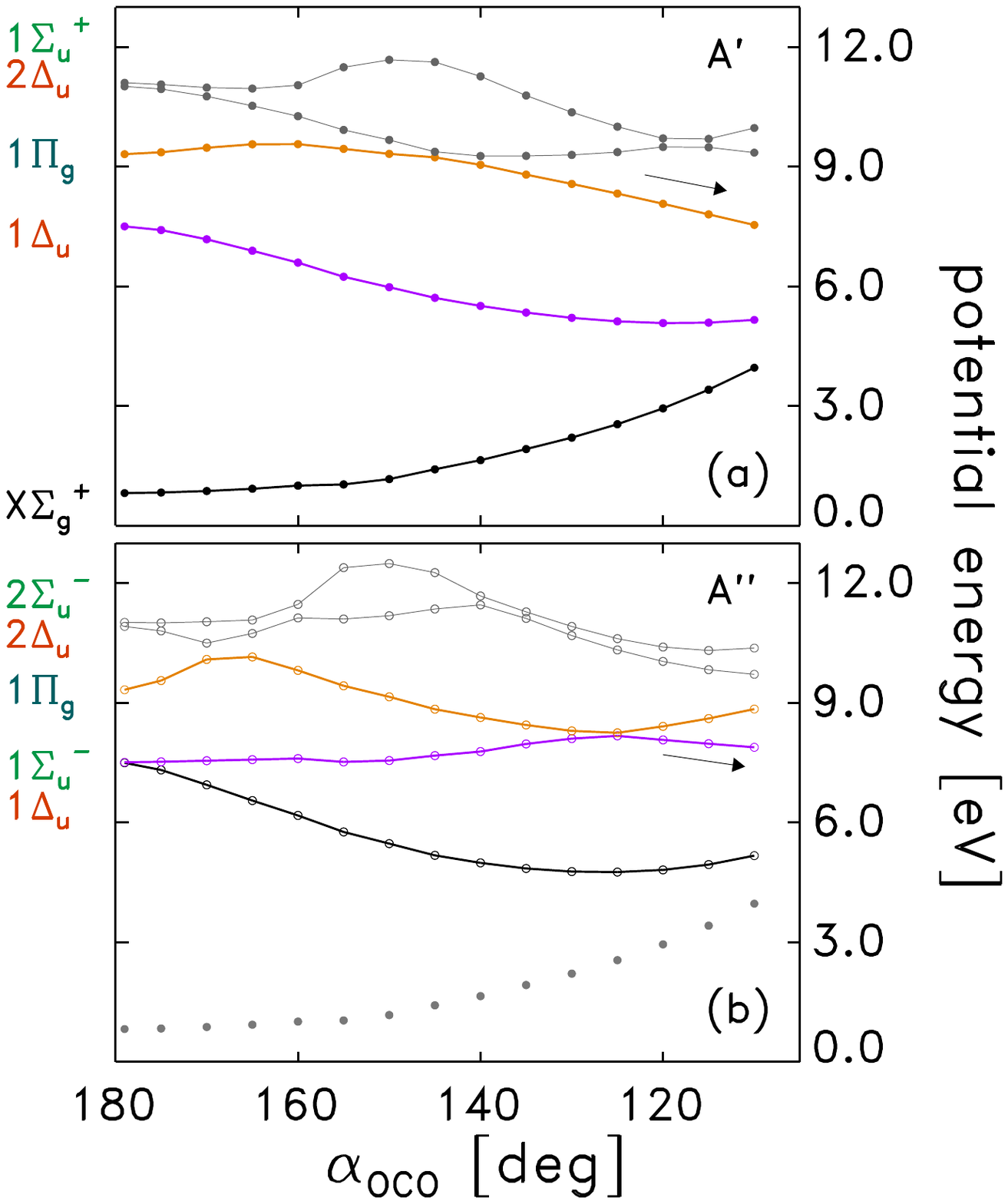}

\vspace{-1cm}

Fig.\ 12

\newpage
\mbox{ }
\vspace{-2cm}

\includegraphics[angle=0,scale=0.7]{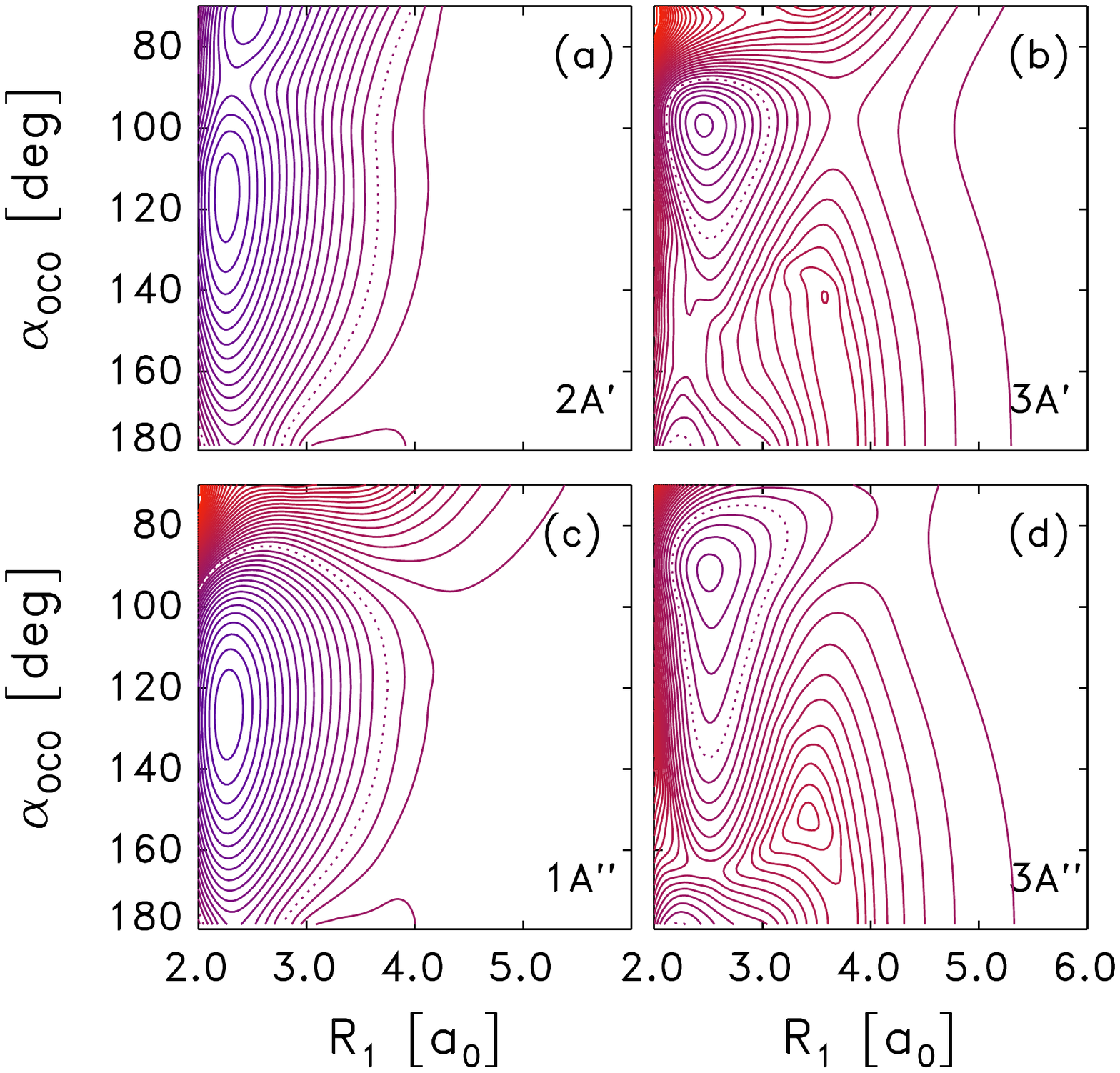}

\vspace{-1cm}

Fig.\ 13

\newpage
\mbox{ }
\vspace{1cm}

\includegraphics[angle=0,scale=0.7]{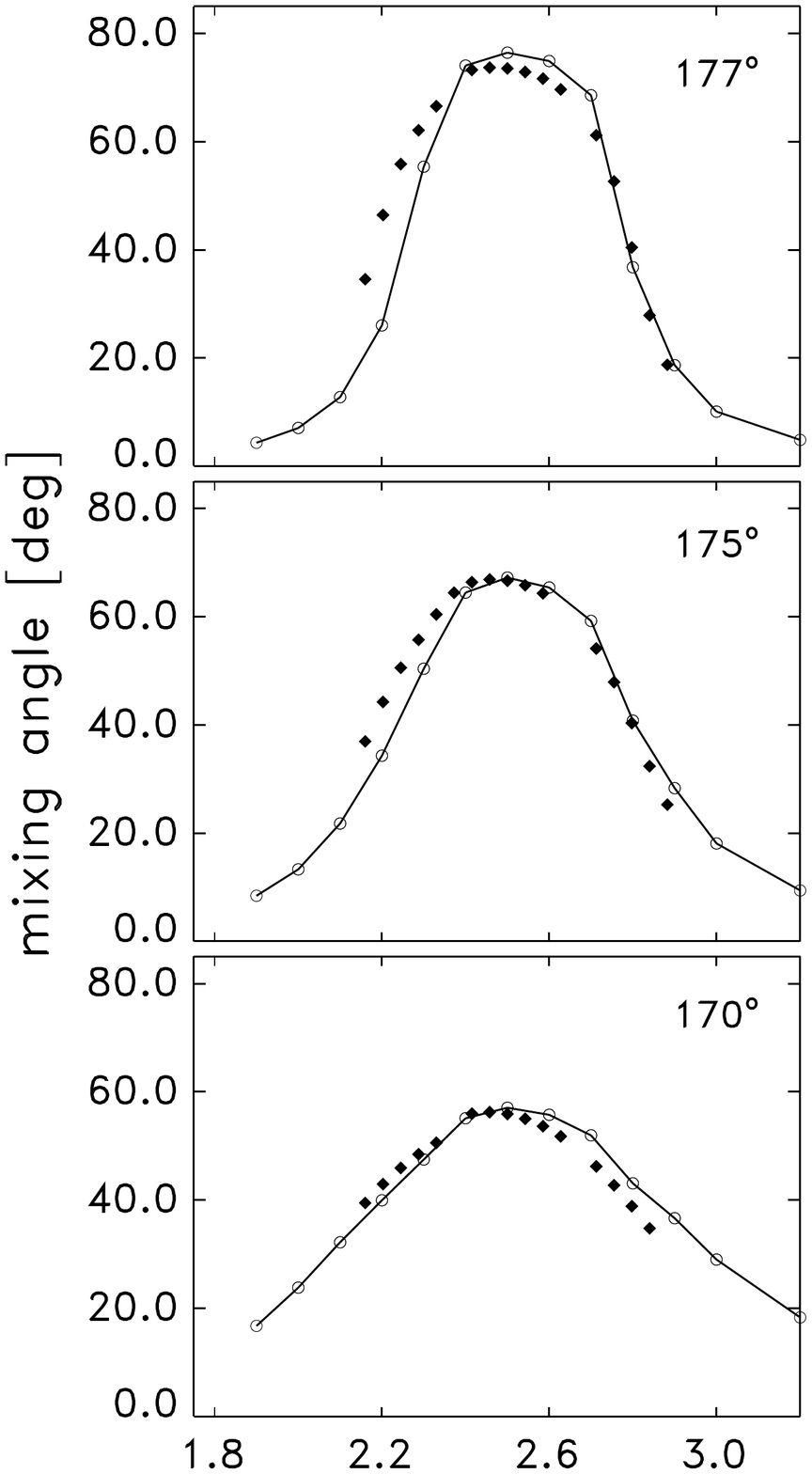}

\vspace{-1cm}

Fig.\ 14

\newpage
\mbox{ }
\vspace{1cm}

\includegraphics[angle=0,scale=0.7]{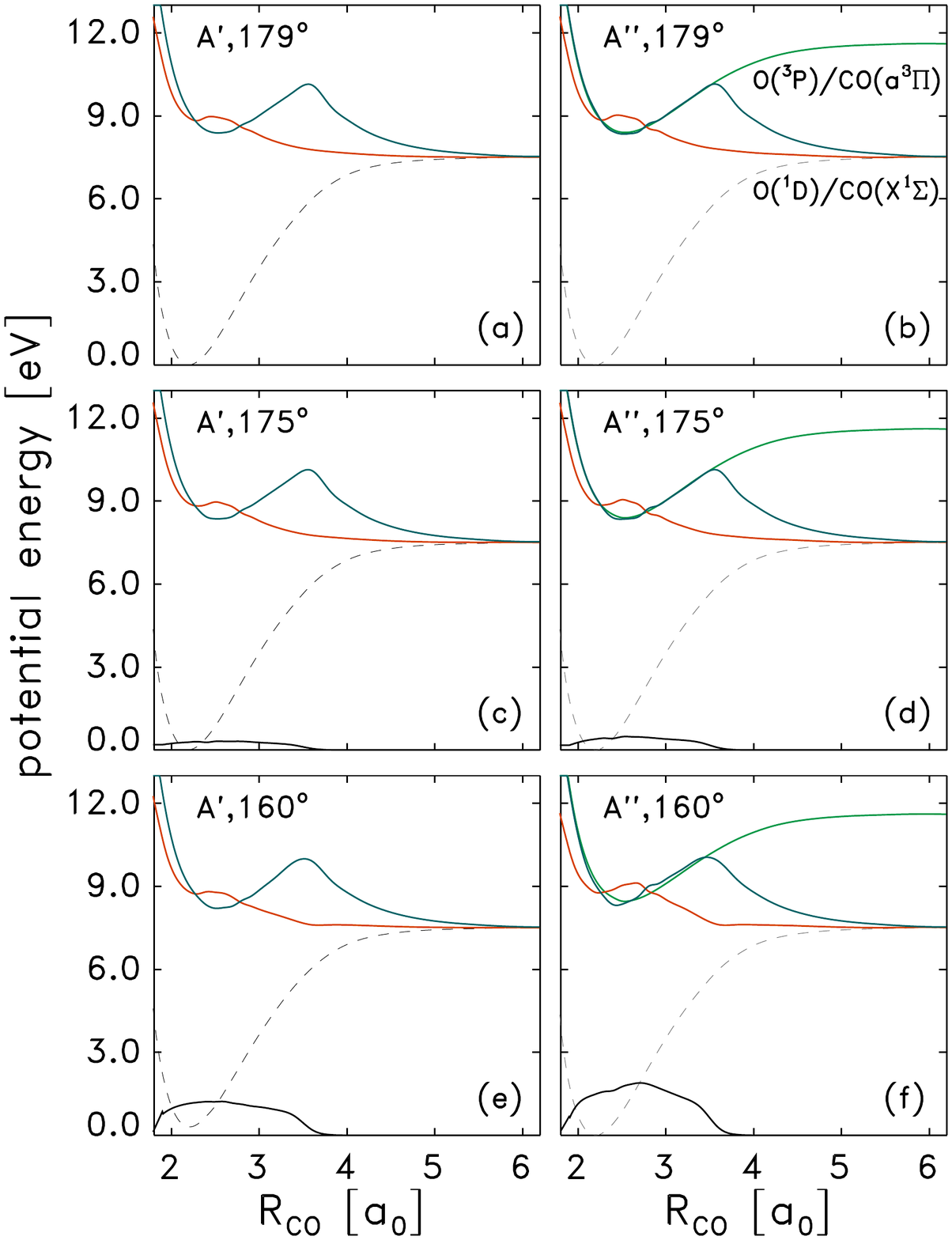}

\vspace{-1cm}

Fig.\ 15

\newpage
\mbox{ }
\vspace{-1cm}

\includegraphics[angle=0,scale=0.7]{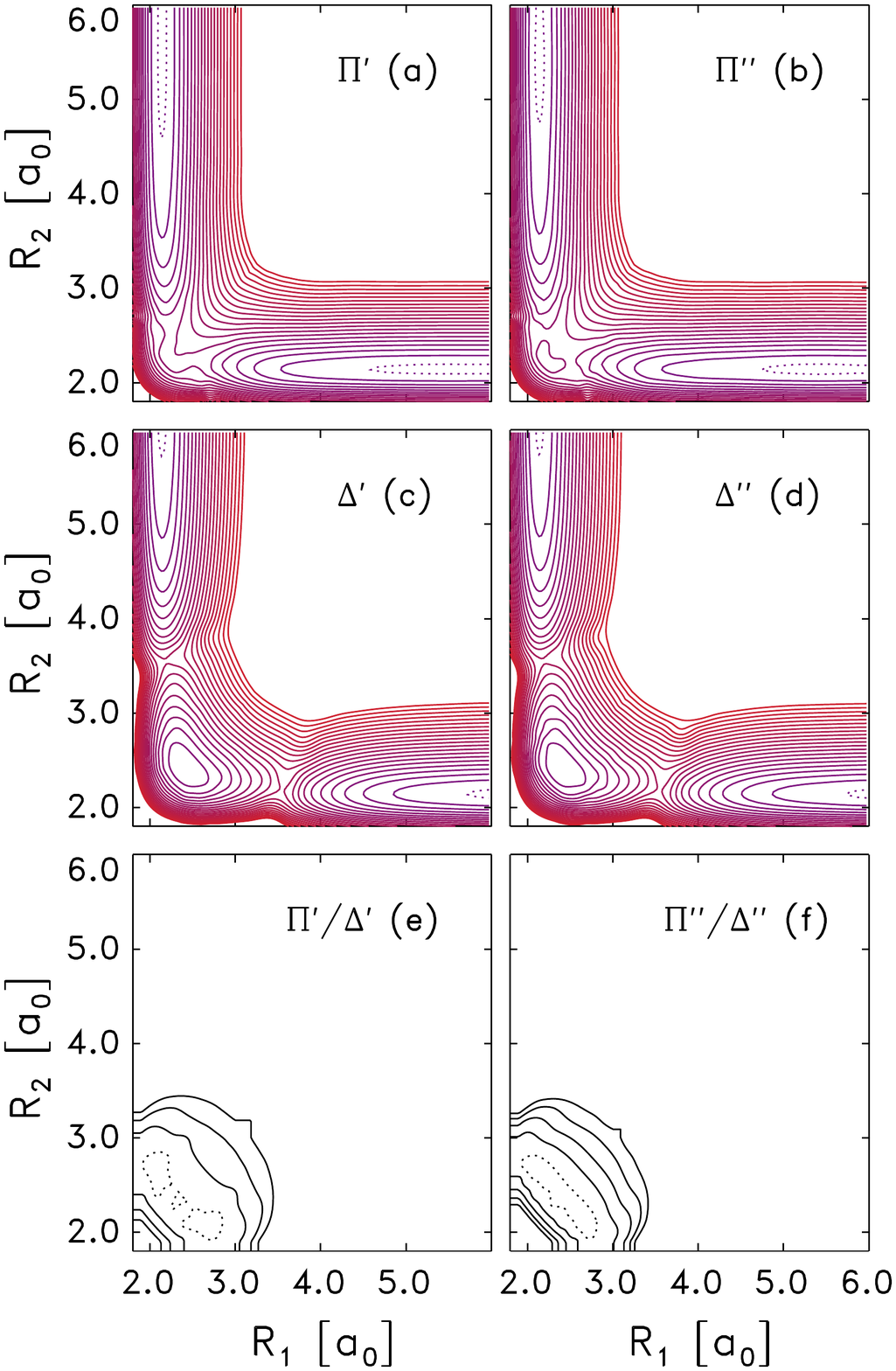}

\vspace{-1cm}

Fig.\ 16

\newpage
\mbox{ }
\vspace{-1cm}

\includegraphics[angle=0,scale=0.7]{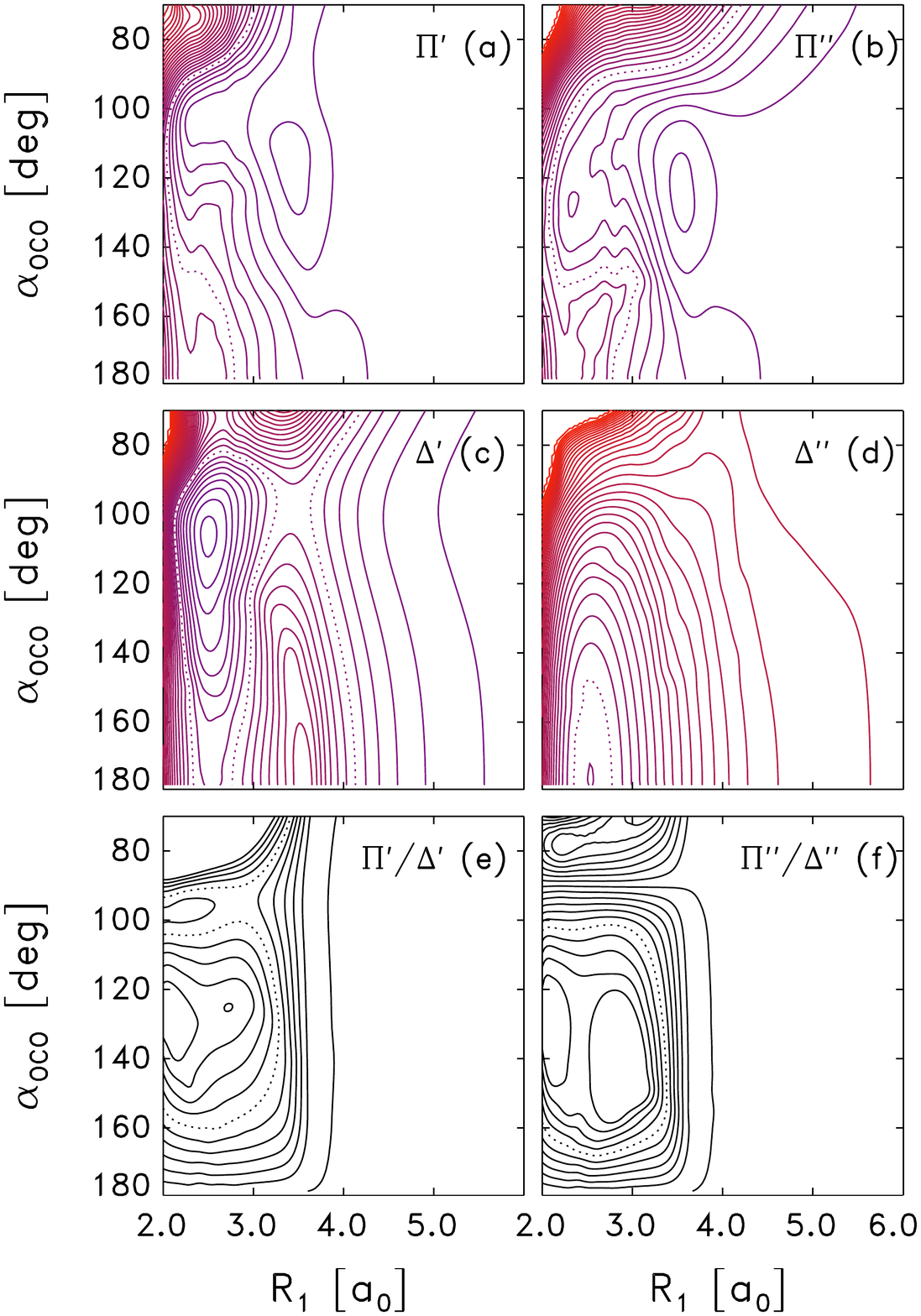}

\vspace{-1cm}

Fig.\ 17

\end{document}